\documentclass[dvipdfmx,hiresbb,onecolumn]{pasj01}

\Received{yyyy/mm/dd}
\Accepted{yyyy/mm/dd}

\usepackage{lineno}

\newcommand{\argmax}{\mathop{\rm arg~max}\limits}
\newcommand{\argmin}{\mathop{\rm arg~min}\limits}

\begin{document} 

\title{ 
  Hitomi-HXT deconvolution imaging of the Crab Nebula
  dazzled by the Crab pulsar}
%
\author{Mikio \textsc{Morii}\altaffilmark{1,2}%
}
\altaffiltext{1}{Institute of Space and Astronautical Science,
  Japan Aerospace Exploration Agency,
  3-1-1 Yoshinodai, Chuo-ku, Sagamihara, Kanagawa 252-5210, Japan}
\altaffiltext{2}{DATUM STUDIO Company Limited,
  Toranomon Hills Business Tower 27F,
  1-17-1 Toranomon, Minato-ku, Tokyo 105-6427, Japan}
\email{morii.mikio@jaxa.jp}
\author{Yoshitomo \textsc{Maeda}\altaffilmark{1}}
\author{Hisamitsu \textsc{Awaki}\altaffilmark{3}}
\altaffiltext{3}{Department of Physics, Ehime University,
  2-5 Bunkyou-cho, Matsuyama, Ehime 790-8577, Japan}
\author{Kouichi \textsc{Hagino}\altaffilmark{4}}
\altaffiltext{4}{Department of Physics, The University of Tokyo,
  7-3-1 Hongou, Bunkyo-ku, Tokyo 113-0033, Japan}
\author{Manabu \textsc{Ishida}\altaffilmark{1}}
\and
\author{Koji \textsc{Mori}\altaffilmark{5}}
\altaffiltext{5}{Department of Applied Physics and Electronic Engineering,
  University of Miyazaki, Miyazaki 889-2192, Japan}



\maketitle

\begin{abstract}
  We develop a new deconvolution method to improve the
  angular resolution of the Crab Nebula image
  taken by the {\it Hitomi} HXT.
  Here, we extend the Richardson-Lucy method by introducing
  two components for the nebula and the Crab pulsar
  with regularization for smoothness and flux, respectively,
  and deconvolving multi-pulse-phase images simultaneously.
  The deconvolved nebular image at the lowest energy band of 3.6--15 keV
  looks consistent with the {\it Chandra} X-ray image.
  Above 15 keV, we confirm that the {\it NuSTAR}'s findings
  that the nebula size decreases in higher energy bands.
  We find that the north-east side of the nebula becomes
  dark in higher energy bands.
  Our deconvolution method can be applicable
  for any telescope images of faint diffuse objects
  containing a bright point source.
\end{abstract}


\section{Introduction}

The Crab Nebula is a synchrotron nebula powered by the rotational energy loss
of the central pulsating neutron star, the Crab pulsar
\citep{Rees_Gunn_1974, Kennel_Coroniti_1984}.
The X-ray image of the nebula with a high angular resolution
of arcsec level was obtained by the {\it Chandra X-Ray Observatory}
below about 10 keV ($0.5$ arcsec; \cite{Weisskopf+2000}).
It revealed the detailed nebular structure around the Crab pulsar such as
torus, inner ring and jets.
The torus whose symmetric axis coincides with the pulsar spin axis
is seen and the north-west edge is closer to the observer.
The jets emerge from the Crab pulsar along the pulsar spin axis
outwards into the counter directions of south-east and north-west.
Such structures elucidated the magneto-hydrodynamics and
particle acceleration in the pulsar wind nebula
(e.g., \cite{Porth_Komissarov_Keppens_2014}).
The energy dependency of the shape would help
to understand the particle acceleration mechanism within the pulsar wind.
Above 10 keV, however, the angular resolution is worse
than that below 10 keV and the best resolution so far is
$18$~arcsec (full width at half maximum; FWHM)
and $58$~arcsec (half power diameter; HPD),
achieved by the {\it NuSTAR} X-ray telescope (\cite{Harrison+2013}).

The {\it Hitomi} X-ray Observatory was launched in February 2016
and stopped its operation on the end of March
(\cite{Takahashi+2016}).
It carried the hard X-ray imaging spectroscopy system
consisting of two pairs of Hard X-ray Imagers (HXI-1 and HXI-2)
and Hard X-ray Telescopes (HXT-1 and HXT-2).
The HXIs provide images and spectra up to 80 keV
with moderate energy resolution
(1.0 keV at 13.9 keV and 2.0 keV at 59.5 keV in FWHM;
\cite{Hagino+2018}).
The effective area is about 300 cm$^2$ at 30 keV
with a field of view of 9 $\times$ 9 arcmin$^2$.
The angular resolution of the {\it Hitomi} HXT was reported to be
$\sim 1.6$ arcmin in HPD \citep{Matsumoto+2018}.
Although it is slightly worse than that of the {\it NuSTAR},
we report in this paper that
the core of the point spread function (PSF) of the {\it Hitomi} HXT
we use is smaller than that of the {\it NuSTAR}
(section \ref{sec: fwhm}).
So, {\it Hitomi} HXT has a potential to obtain
better images in angular resolution than that of {\it NuSTAR}
after image deconvolution.

Image deconvolution is a mathematical technique to improve
the angular resolution of telescopes, which utilizes information of
the PSF of telescopes.
Richardson-Lucy method \citep{Richardson_1972, Lucy_1974}
is a well-known canonical method for the image
deconvolution, which is an iterative computational algorithm
to obtain the deconvolved image of the sky
with maximal likelihood for the image data,
assuming that the detected photon events follow the Poisson distribution.
Since the Crab Nebula is a spatially extended and diffuse object,
the smoothness in the intensity distribution
is a well convincing assumption and useful to reduce
the statistical fluctuation.
So, the image deconvolution with smoothness and sparseness regularization
proposed by \citet{Morii+2019} is thought to be applicable
for this case.
However, the smoothness regularization make the nebula image deteriorate,
because the far bright Crab pulsar overlays the nebula image.
It motivates us to develop a new deconvolution method
special for the Crab Nebula,
separating the nebula and pulsar components effectively.

\section{Observation data}

The Crab Nebula was observed on 2016 March 25 from 12:35 to 18:01 UT
during the {\it Hitomi}'s commissioning phase.
It was imaged at around the center of the HXI array.
The observation time span was 21.5~ks,
whereas the total on-source time was $\sim 8$~ks.
It is the only observation of the Crab Nebula taken by {\it Hitomi}.
We use the cleaned data (Sequence ID. 100044010
\footnote{It is available at the web sites of
HEASARC https://heasarc.gsfc.nasa.gov/docs/hitomi/archive/
and DARTS https://www.darts.isas.jaxa.jp/astro/hitomi/.}),
made by applying the standard screening \citep{Angelini+2016}
with the processing script of the version of 01.01.003.003.
To detect the pulsation of the Crab pulsar,
the barycentric correction for photon arrival times
is applied for the Crab pulsar position of
$(\alpha,\delta)_{\rm J2000} = (83.6332208,22.0144614)$
(\cite{Mori+2004} and reference therein).

We use all the cleaned data for HXI-1,
whereas we use only the hard band data above 15 keV for HXT-2.
It is because one of HXI-2's readout strips near the aimpoint
was a bit noisy, making a bad line in the image,
and has no sensitivity below $\sim 10$~keV.
In what follows, we divide the HXI data into three energy bands:
3.6--15, 15--30 and 30--70 keV.
The total counts within the square region with $80 \times 80$ pixels
centered on the Crab pulsar are summarized in table~\ref{tab: hxt_counts}.
The images in the lowest energy band of 3.6--15 keV have
rich sample of photons over million counts,
whereas the highest energy band images
above 30 keV have less sample of photons
of forty thousands.

\begin{table}
  \caption{Counts detected in the energy bands of HXIs}
  \begin{center}
    \begin{tabular}{lrrrrrrc} \hline
      Detector & \multicolumn{3}{c}{HXI-1} &
      \multicolumn{3}{c}{HXI-2} & Note \\ \hline
      & \multicolumn{3}{c}{Energy band (keV--keV)}
      & \multicolumn{3}{c}{Energy band (keV--keV)} & \\
      Phase ID & 3.6--15 & 15--30 & 30--70 &
      3.6--15 & 15--30 & 30--70 & \\ \hline
      ALL  & 1 696 836  & 223 323 & 43 878 &
      1 545 317  & 204 465 & 39 940 & \\
      ON2  &   462 350  &  62 520 & 12 564 &
      418 696  &  57 225 & 11 490 & Secondary peak \\
      OFF1 &   613 895  &  78 346 & 14 953 &
      562 049  &  71 508 & 13 649 & Off-pulse \\
      ON1  &   375 538  &  50 197 &  9 926 &
      341 238  &  45 754 &  8 817 & Primary pulse \\
      OFF2 &   245 053  &  32 260 &  6 435 &
      223 334  &  29 978 &  5 984 & \\
      \hline
    \end{tabular}
  \end{center}
  \label{tab: hxt_counts}
\end{table}

In order to estimate the non X-ray background level,
we analyze a blank sky data
since the non X-ray background dominates the cosmic X-ray background
in flux \citep{Hagino+2018}.
As listed in \citet{Hagino+2018}, RX J1856.5$-$3754 has
no significant flux above 2 keV and is regarded
as a blank sky for our purpose.
We combine the cleaned RX J1856.5$-$3754 data
with the sequence IDs of 100043010, 100043020, 100043030, 100043040
and 100043050, and make images of 80 $\times$ 80 pixels
in the sky coordinate for each energy band.
The total exposure time of RX J1856.5$-$3754 was 102~ks for HXIs.
The estimated counts of non X-ray background per 8~ks
are 11(16), 9(10) and 40(40) in the energy bands
of $3.6-15$, $15-30$ and $30-70$ keV for HXI-1 (HXI-2),
respectively. They are negligibly small for our purpose.

\section{Point spread function}

\begin{figure}
  \begin{center}
    \includegraphics[width=16.0cm]{
      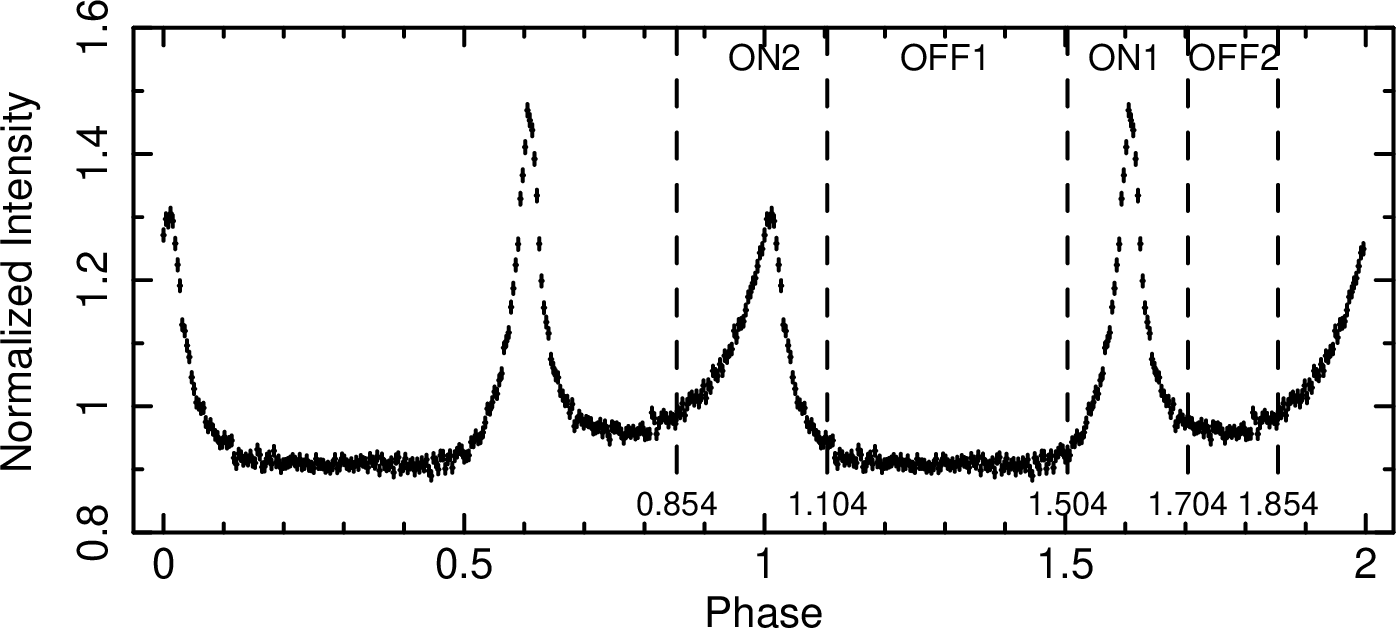}
  \end{center}
  \caption{Folded light curve of the HXI-1 in all the energy band.}
  \label{fig: hxt1_foldedlightcurve}
\end{figure}

For the image deconvolution,
the model of the PSF dependent on energy and in-coming directions
of photons (response matrix) is necessary.
Due to the {\it Hitomi}’s short life,
the PSF is not well modeled and the dependency on the direction
is also not measured.
However, at least for the Crab Nebula,
a reliable PSF is obtained for the in-coming direction
from the Crab pulsar
as demonstrated by \citet{Matsumoto+2018}.
We here follow their method to make the PSF.

The Crab pulsar exhibits X-ray pulsation with a period of $\sim 34$ ms,
and the pulse profile has a double peak structure \citep{Ducros+1970}.
Since the HXI has a time resolution
of $25.6~\mu{\rm s}$ \citep{Nakazawa+2018},
the pulsation was successfully detected
by the HXI as reported by \citet{Hitomi_Collaboration+2018}.
Figure \ref{fig: hxt1_foldedlightcurve} shows such a pulse profile
of the Crab pulsar, made by folding the HXI-1 light curve
with the pulse period of 33.7204626 ms at 57472.5~d (MJD),
which is the phase zero.
We define the pulse phases ON1, OFF2, ON2 and OFF1
by the durations of $0.504-0.704$, $0.704-0.854$,
$0.854-1.104$ and $0.104-0.504$, respectively.
Table \ref{tab: hxt_counts} shows the counts in each pulse phase.
Then, by subtracting the image in the OFF1 phase
from the ON phase (the addition of the ON1 and ON2 phases),
we can obtain the Crab pulsar image,
which is just the PSF we need for the image deconvolution.
We use the square region with $80 \times 80$ pixels
centered on the Crab pulsar for the PSF.

Since this PSF is obtained from the observed data itself,
the pointing instability of the telescope during the observation
does not cause the uncertainty of the PSF.
We adapt the same PSF for the directions
other than the central pulsar direction.
In section~\ref{sec: uncertainty},
we discuss the causes of uncertainty of the PSF
and evaluate the effect of them for the deconvolved image.

\section{Sharpness of the PSF}
\label{sec: fwhm}

\begin{figure}
  \begin{center}
    \includegraphics[width=10.0cm]{
      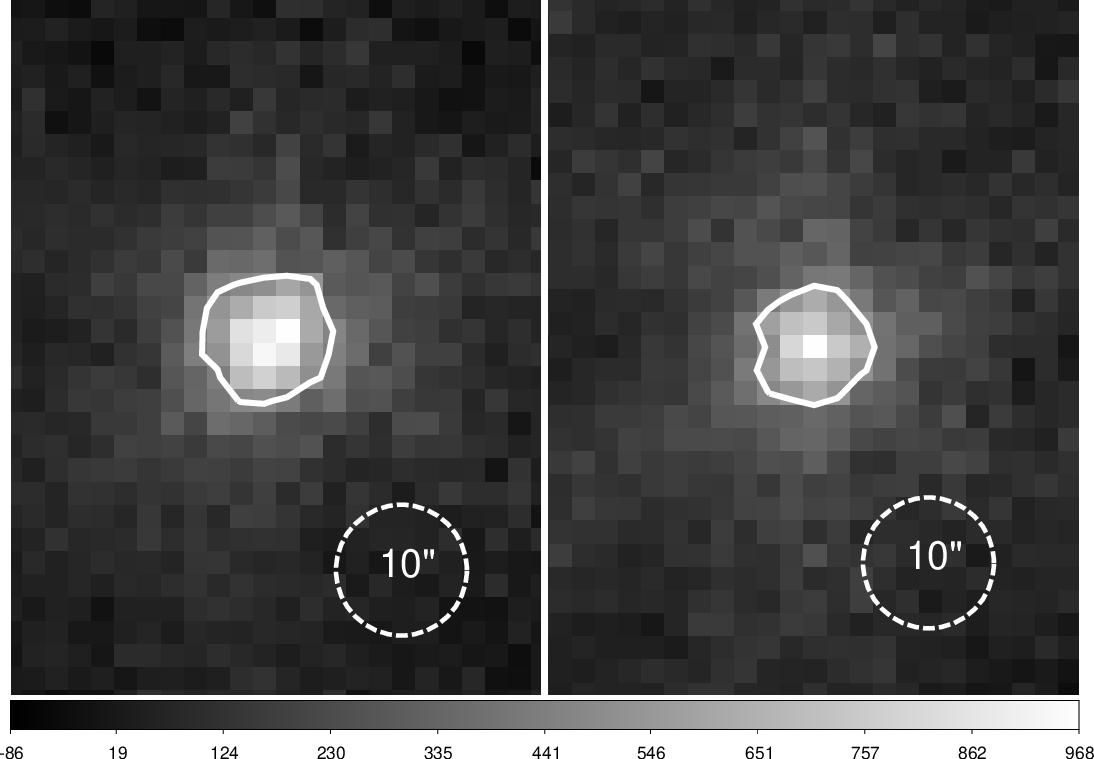}
  \end{center}
  \caption{The Crab pulsar images of HXI-1 (left) and HXI-2 (right).
    They are made by subtracting the OFF-pulse image from
    the ON-pulse image (see text for details).
    The pixel size of the image is 1.77 arcsecs.
    The contours are drawn at the half of each peak counts.
    The diameters of the dashed circles are 10 arcsecs.}
  \label{fig: fwhm}
\end{figure}

\begin{table}
  \caption{FWHM of the Crab pulsar images}
  \begin{center}
    \begin{tabular}{ccccc} \hline
      & \multicolumn{2}{c}{HXI-1}  & \multicolumn{2}{c}{HXI-2} \\
      \hline
      Energy band & FWHM & Maximum pixel & FWHM & Maximum pixel \\
      (keV--keV) & (arcsec) & (counts/pixel) & (arcsec)
      & (counts/pixel) \\ \hline
      All band   & 10 & 1012 &  9 & 967 \\
      3.6-15    & 10 & 852  &  8 & 813\\
      15-30     & 10 & 138  & 10 & 131 \\
      30-70     &  5 & 43   &  7 & 34 \\  \hline
    \end{tabular}
  \end{center}
  \label{tab: fwhm}
\end{table}

Since the sharpness of the PSF core contributes
the goodness of an angular resolution of deconvolved images,
we here evaluate it as the FWHM of the PSF.
Figure \ref{fig: fwhm} shows the Crab pulsar images of HXIs
without smoothing nor binning,
for which no energy filtering are applied.
The counts at the peak pixels are 1012 and 967
for HXI-1 and HXI-2, respectively.
The contours depict the position of the half of the peaks.
Before drawing these contours,
we applied a Gaussian smoothing with one sigma width of one pixel,
to draw the contours in sub-pixel resolution.
We calculate the FWHM as the equivalent diameter of the circle
whose area is the same as the enclosed area by the contour.
The FWHMs of the PSFs are 10 and 9 arcsecs for HXI-1 and HXI-2,
respectively. 
They are about a half of the PSF of the {\it NuSTAR} telescope
(18 arcsecs : \cite{Harrison+2013}).
The FWHMs of the PSFs for three energy bands
are shown in table \ref{tab: fwhm}.

\section{Mathematical formulation of the problem}

{\it Hitomi} satellite has two hard X-ray telescopes (HXT-1, HXT-2)
with photon counters (HXI-1, HXI-2) at their respective foci.
We consider a rectangular area on the tangential plane
of the celestial sphere as an observed image
by the {\it Hitomi} HXT/HXI system.
The area is divided into $M = m \times n$ pixels ($m = 100, n = 100$),
where each pixel is indexed with
$u = (i, j)$ ($i = 1, \cdots, m; j = 1, \cdots, n$).
The image is expressed by the distribution of
intensity of photons $I_u$ per each pixel $u$, by a non-negative real value.
Pixels of HXI is indexed with $v$ ($v = 1, \cdots, V$; $V = 80 \times 80$).
The HXI-1 and HXI-2 are identified with $c = 1, 2$, respectively,
and the number of counters is $n_c = 2$.
The events detected at a pixel $v$ of HXI ($c$)
consist of both X-ray photons through {\it Hitomi}-HXT/HXI system and
charged particles which directly hit on the pixel.
The number of these events in an exposure (an observation)
of {\it Hitomi}-HXT/HXI is $Y_{c,v}$, which is a non-negative integer value.
It follows a Poisson distribution:
\begin{equation}
  Y_{c,v} \sim {\rm Poisson} \left( \sum_u t_{c,v,u} I_u + b_{c,v} \right),
  \label{eq: Y_cv}
\end{equation}
where $t_{c,v,u}$ is the response of the {\it Hitomi}-HXT/HXI.
It means a distribution of the expected photon counts at pixel $v$ of
counter $c$ from a pixel $u$ with a unit intensity
at the celestial sphere, that is the PSF.
Here, $b_{c,v}$ is the particle background counts
at pixel $v$ of the counter $c$
as a non-negative real value.

The detection efficiency ($\epsilon_u$) of photons by the counters
from a celestial pixel $u$ follows a relation
$\sum_{c,v} t_{c,v,u} = \epsilon_u \le 1$.
Here, the effects of vignetting of HXTs and quantum efficiency
of HXIs are included in the factor $\epsilon_u$.
Assuming $0 < \epsilon_u \le 1$, and setting
$t^\prime_{c,v,u} = t_{c,v,u} / \epsilon_u$ and
$I^\prime_u = \epsilon_u I_u$,
the equation (\ref{eq: Y_cv}) becomes
\begin{equation}
  Y_{c,v} \sim {\rm Poisson} \left(
    \sum_u t^\prime_{c,v,u} I^\prime_u + b_{c,v} \right),
  \label{eq: Y_cv_prime}
\end{equation}
and $t^\prime_{c,v,u}$ follows $\sum_{c,v} t^\prime_{c,v,u} = 1$.
So, we solve for $I^\prime_u$ and $I_u$ is then obtained by
$I_u = I_u^\prime / \epsilon_u$.
Hereafter, we thus replace $t^\prime_{c,v,u} \rightarrow t_{c,v,u}$
and $I_u^\prime \rightarrow I_u$,
so $\sum_{c,v} t_{c,v,u} = 1$ is satisfied.

\section{Case of the Crab Nebula with the Crab pulsar}

We derive the mathematical formulation
for the case of the Crab Nebula with the Crab pulsar.
Remarks of this case are the followings:
(1) The Crab pulsar is a bright point source locating at the center of
the Crab Nebula. Its intensity is concentrated at one pixel
and far brighter than those of the surrounding nebula.
(2) Intensity of the Crab pulsar changes periodically in the pulse
period of $\sim 34$~ms with large variation.
Due to the first one, we have to assume two components
for the celestial image. To decouple these intensities at the
pulsar position, we assume spatial smoothness of the nebula component.
Utilizing the second point,
decoupling of two components is clarified further by introducing
a simultaneous image deconvolution using the images of
all pulse phases at once,
with the nebula component common in all phases
and only the intensity of the pulsar varied.

We divide observed data into $n_p$ pulse phases ($n_p = 10$),
then we obtain $n_p$ images observed by the {\it Hitomi} HXT.
Pulse phases are indexed with $p = 1, 2, \cdots, n_p$,
and the ratio of the time width of the phase $p$ to the pulse period
is $(\Delta \phi)_p$,
where $0 < (\Delta \phi)_p \le 1$ and
$\sum_{p=1}^{n_p} (\Delta \phi)_p = 1$.
In the case of equivalent time width for phases,
$(\Delta \phi)_p = 1 / n_p$.
We express the image of the Crab Nebula by $I_u$
and the Crab pulsar by $I_{0,u}$.
Here, the value of $I_{0,u}$ is one at the position of the Crab pulsar,
and zero at the other positions.
The intensity of the Crab pulsar at phase $p$ is expressed by $f_p$.

\begin{table}
  \caption{Livetime fraction in each pulse phase}
  \begin{center}
    \begin{tabular}{lcc} \hline
      Phase & HXI-1& HXI-2 \\ \hline
      0.0-0.1 &      0.751 & 0.736 \\
      0.1-0.2 &      0.782 & 0.768 \\
      0.2-0.3 &      0.784 & 0.770 \\
      0.3-0.4 &      0.784 & 0.770 \\
      0.4-0.5 &      0.783 & 0.769 \\
      0.5-0.6 &      0.754 & 0.737 \\
      0.6-0.7 &      0.741 & 0.726 \\
      0.7-0.8 &      0.774 & 0.759 \\
      0.8-0.9 &      0.768 & 0.754 \\
      0.9-1.0 &      0.742 & 0.726 \\
      \hline
    \end{tabular}
  \end{center}
  \label{tab: hxt_deadtime2}
\end{table}

Due to the limitation of the HXI capability,
HXIs have deadtime in an observing duration, which tends to
be larger for higher event rates. Thus, deadtimes vary in pulse phases.
The livetime is the complement of the deadtime during an observing duration.
Table \ref{tab: hxt_deadtime2} shows the livetime fractions of
observing data of the Crab Nebula, calculated
by the method given in the appendix of \citet{Matsumoto+2018}.
We set the fraction of the livetime to an exposure of pulse phase $p$
for HXI $c$ to be $(F_{\rm lt})_{c,p}$ ($0 \le (F_{\rm lt})_{c,p} \le 1$).

The number of events in the pixel $v$ of the counter $c$
in an exposure of a pulse phase $p$ 
is $Y_{c,p,v}$, which is a non-negative integer value.
It follows a Poisson distribution:
\begin{equation}
  Y_{c,p,v} \sim {\rm Poisson} \left\{ \left[
  \sum_u t_{c,v,u}(I_u + f_p I_{0,u}) + b_{c,v} \right]
    (\Delta \phi)_p (F_{\rm lt})_{c,p} \right\}.
  \label{eq: Y_cpv}
\end{equation}
Thus, the likelihood of detecting $Y$ events given $I$ and $f$
is expressed as follows:
\begin{equation}
  p(Y|I,f) = \prod_{c=1}^{n_c} \prod_{p=1}^{n_p} \prod_{v=1}^V
  {\rm Poisson} \left\{ Y_{c,p,v};
  \left[\sum_u t_{c,v,u}(I_u + f_p I_{0,u}) + b_{c,v} \right]
  (\Delta \phi)_p (F_{\rm lt})_{c,p} \right\}.
\end{equation}

\section{Regularization}

We introduce a smoothness constraint for the nebula component $I_u$
to reduce statistical fluctuation and to decouple the nebula and
pulsar components at the pulsar position.
In addition, to stabilize the nebula image around the pulsar position,
we also introduce the flux constraint for the pulsar flux $f_p$.
Such constraints are expressed by prior probabilities in Bayesian inference.
The prior for $I_u$ is
\begin{equation}
  p_{\rm smooth}(I) = Z_I \exp[-\mu V(I)],
\end{equation}
where
\begin{eqnarray}
  V(I) & = & \sum_{(r,s)\in N} (I_r - I_s)^2 \nonumber \\
  & = & \sum_{i = 1}^{m - 1} \sum_{j = 1}^{n - 1}
  \left[ (I_{i,j} - I_{i + 1,j})^2
    + (I_{i,j} - I_{i,j + 1})^2 
    \right] \nonumber \\
  & & + \sum_{i = 1}^{m - 1} (I_{i,n} - I_{i + 1,n})^2
  + \sum_{j = 1}^{n - 1} (I_{m,j} - I_{m, j+1})^2 .
\end{eqnarray}
Here, $Z_I$ is a normalization constant.
$\sum_{(r,s)\in N}$ denotes the summation between two adjoining pixels.
$\mu > 0$ is a hyper-parameter to control the smoothness
of the nebula image.

The prior for $f_p$ is
\begin{equation}
  p_{\rm flux}(f) = Z_f \exp[-\gamma D(f)],
\end{equation}
where
\begin{equation}
  D(f) = \sum_p (f_p - f_{0,p})^2 .
\end{equation}
Here, $Z_f$ is a normalization constant.
$\gamma > 0$ is a hyper-parameter to make the flux
be close to the pulsed flux of the Crab pulsar $f_{0,p}$.
Because {\it Hitomi} HXT cannot resolve the Crab pulsar and
the nebula on the image,
we here use the pulsed flux instead of the unknown total flux.
The $\gamma$ parameter works to adjust the difference
between the pulsed and total fluxes of the Crab pulsar.
The pulsed flux $f_{0,p}$ is obtained by
\begin{equation}
  f_{0,p} = \frac{1}{(\Delta\phi)_p}
  \sum_c \frac{\sum_v Y_{c,p,v}}{(F_{\rm lt})_{c,p}}
  - \frac{1}{(\Delta\phi)_{\rm off}}
  \sum_c \frac{\sum_v Y_{c,{\rm off},v}}{(F_{\rm lt})_{c,{\rm off}}},
\end{equation}
where off is the pulse minimum phase.

Then, the posterior is obtained by
\begin{eqnarray}
  p(I,f|Y) & = & \frac{p(I,f,Y)}{p(Y)} =
  \frac{p_{\rm smooth}(I) p_{\rm flux}(f) p(Y|I,f)}{p(Y)}.
\end{eqnarray}
We solve $I$ and $f$ by maximizing
the logarithm of $p(I,f|Y)$ ($= - L_{\rm cost}(I,f)$):
\begin{eqnarray}
  \log p(I,f|Y) & = & \log p(Y|I,f) + \log p_{\rm smooth}(I)
  + \log p_{\rm flux}(f) + {\rm const.} \nonumber \\
  & = & \log p(Y|I,f) - \mu V(I) - \gamma D(f) + {\rm const}.
\end{eqnarray}

\section{Optimization of the likelihood part $p(Y|I,f)$}

We use the EM (Expectation-Maximization) algorithm
\citep{Bishop_06, Dempster+1977}
to maximize the likelihood part $p(Y|I,f)$.
In this frame work, $p(Y|I,f)$ is called
observed likelihood. The corresponding
complete likelihood is expressed by
\begin{eqnarray}
  p_{\rm complete}(Y,z,w,e|I,f) & = &
  \prod_{c=1}^{n_c} \prod_{p=1}^{n_p} \prod_{v=1}^V \left\{
  \prod_{u=1}^M{\rm Poisson}\left[z_{c,p,v,u};
    t_{c,v,u}I_u(\Delta\phi)_p(F_{\rm lt})_{c,p}\right]\right.
  \nonumber \\
  & & \times {\rm Poisson}\left[w_{c,p,v};
    f_pd_{0,c,v}(\Delta\phi)_p(F_{\rm lt})_{c,p}\right]
  \nonumber \\
  & & \times {\rm Poisson}\left[
    e_{c,p,v};b_{c,v}(\Delta\phi)_p(F_{\rm lt})_{c,p}\right]
  \nonumber \\
  & & \left.\times
  \chi_{\{Y_{c,p,v}=\sum_u z_{c,p,v,u}+w_{c,p,v}+e_{c,p,v}\}}(z,w,e)\right\},
\end{eqnarray}
where $z_{c,p,v,u}$, $w_{c,p,v}$ and $e_{c,p,v}$ are latent variables.
$z_{c,p,v,u}$ is the photon counts in the pulse phase $p$
at the pixel $v$ of the detector $c$ through the {\it Hitomi}-HXT/HXI system
from the celestial pixel $u$ of the nebula component.
$w_{c,p,v}$ is the same photon counts from the pulsar component.
$e_{c,p,v}$ is the counts of charged particles
in the pulse phase $p$ at the pixel $v$ of the detector $c$.
$d_{0,c,v} = \sum_u t_{c,v,u} I_{0,u}$ is the image
of the Crab pulsar on the counter $c$.
$\chi_A$ is the indicator function of a set $A$.
Indeed, $p_{\rm complete}(Y,z,w,e|I,f)$ is the complete likelihood,
because it satisfies the following:
\begin{equation}
  p_{\rm observed}(Y|I,f)
  = \sum_{z,w,e} p_{\rm complete}(Y,z,w,e|I,f).
\end{equation}
So, the $r$-th iterative step of the EM algorithm is derived to be
\begin{eqnarray}
  (I,f)^{(r+1)} & = &
  \argmax_{(I,f)} {\cal L}\left[p(z,w,e|Y,I^{(r)},f^{(r)}), (I,f)\right]
  \nonumber \\
  & = & \argmax_{(I,f)} \sum_{(z,w,e); p(z,w,e|Y,I^{(r)},f^{(r)}) \neq 0}
  p(z,w,e|Y,I^{(r)},f^{(r)}) \log \frac{p(z,w,e,Y|I,f)}
  {p(z,w,e|Y,I^{(r)},f^{(r)})}
  \nonumber\\
  & = & \argmax_{(I,f)} \sum_{(z,w,e); p(z,w,e|Y,I^{(r)},f^{(r)}) \neq 0}
  p(z,w,e|Y,I^{(r)},f^{(r)}) \log p(z,w,e,Y|I,f) \nonumber\\
  & = & \argmax_{(I,f)} \log p(z^{(r)},w^{(r)},e^{(r)},Y|I,f),
\end{eqnarray}
where
\begin{equation}
  {\cal L}\left[p(z,w,e), (I,f)\right] = \sum_{(z,w,e); p(z,w,e) \neq 0}
  p(z,w,e) \log \frac{p(z,w,e,Y|I,f)}{p(z,w,e)}
\end{equation}
is an infimum function for $\log p_{\rm observed}(Y|I,f)$,
and
\begin{eqnarray}
  z_{c,p,v,u}^{(r)} & = & \sum_{z,w,e}
  p(z,w,e|Y,I^{(r)},f^{(r)}) z_{c,p,v,u} =
  \frac{Y_{c,p,v}t_{c,v,u}I_u^{(r)}}{D_{c,p,v}^{(r)}}
  = Y_{c,p,v}^{\prime (r)} t_{c,v,u}I_u^{(r)}, \\
  w_{c,p,v}^{(r)} & = & \sum_{z,w,e}
  p(z,w,e|Y,I^{(r)},f^{(r)}) w_{c,p,v}
  = \frac{Y_{c,p,v} f_p^{(r)}d_{0,c,v}}{D_{c,p,v}^{(r)}}
  = Y_{c,p,v}^{\prime (r)}f_p^{(r)}d_{0,c,v},\\
  e_{c,p,v}^{(r)} & = & \sum_{z,w,e}
  p(z,w,e|Y,I^{(r)},f^{(r)}) e_{c,p,v}
  = \frac{Y_{c,p,v} b_{c,v}}{D_{c,p,v}^{(r)}}
  = Y_{c,p,v}^{\prime (r)}b_{c,v},
\end{eqnarray}
\begin{eqnarray}
  D_{c,p,v}^{(r)} & = & \sum_{u^\prime} t_{c,v,u^\prime}I_{u^\prime}^{(r)}
  + f_p^{(r)}d_{0,c,v} + b_{c,v},\\
  Y_{c,p,v}^{\prime (r)} & = & \frac{Y_{c,p,v}}{D_{c,p,v}^{(r)}}.
\end{eqnarray}

\section{Optimization of the $V(I)$ part}

We apply the MM (Majorization-Minimization) algorithm
\citep{Hunter_Lange_2000} to optimize the $V(I)$ part
as shown in \citet{Zhou_Alexander_Lange_2011}.
A surrogate function for $V(I)$, that is a supremum for
$V(I)$, is given by
\begin{equation}
  u(I; I^\prime) = \frac{1}{2} \sum_{(r,s)\in N}
  \left\{ \left[2I_r - (I_r^\prime + I_s^\prime)\right]^2
  + \left[2I_s - (I_r^\prime + I_s^\prime)\right]^2 \right\}.
\end{equation}
Here, the relations 
\begin{eqnarray}
  V(I) & = & \sum_{(r,s)\in N} (I_r - I_s)^2
  \le \frac{1}{2} \sum_{(r,s)\in N}
  \left\{ \left[2I_r - (I_r^\prime + I_s^\prime)\right]^2
    + \left[2I_s - (I_r^\prime + I_s^\prime)\right]^2 \right\} \nonumber\\
  & = & u(I; I^\prime)
\end{eqnarray}
and
\begin{equation}
  V(I) = u(I;I)
\end{equation}
hold.
So, this part can be optimized by
the following update:
\begin{equation}
  I^{(t + 1)} = \argmin_{I} u(I;I^{(t)}).
\end{equation}

\section{Optimization of $\log p(I,f|Y)$}

Combining optimization for both the likelihood and prior parts,
the $r$-th updating rule becomes
\begin{eqnarray}
  (I^{(r+1)},f^{(r+1)}) & = & \argmin_{(I,f)} u_{\rm cost}(I,f;I^{(r)},f^{(r)})
  \nonumber\\
  & = & \argmin_{(I,f)} \left\{
  - {\cal L}\left[p(z,w,e|Y,I^{(r)},f^{(r)}), (I,f)\right]
  + \mu u(I;I^{(r)}) + \gamma D(f)\right\}
  \nonumber\\
  & = & \argmin_{(I,f)} \left[
    - \sum_{(z,w,e); p(z,w,e|Y,I^{(r)},f^{(r)})\neq 0}
    p(z,w,e|Y,I^{(r)},f^{(r)})\log p(z,w,e,Y|I,f)\right. \nonumber\\
    & & + \left. \mu u(I;I^{(r)}) + \gamma D(f)\right]\nonumber\\
  & = & \argmin_{(I,f)} \left[ -\sum_{c,p,v} 
    \left(\sum_u\left\{z_{c,p,v,u}^{(r)} \log I_u
    - [t_{c,u,v}I_u(\Delta\phi)_p(F_{\rm lt})_{c,p}] \right\}
    \right.\right.\nonumber\\
    & & \left.\left. + w_{c,p,v}^{(r)} \log f_p
    - [f_p d_{0,c,v}(\Delta\phi)_p(F_{\rm lt})_{c,p}]\right)
    + \mu u(I;I^{(r)}) + \gamma D(f)\right],
\end{eqnarray}
where
\begin{equation}
  u_{\rm cost}(I,f;I^\prime,f^\prime) =
  - {\cal L}\left[p(z,w,e|Y,I^\prime,f^\prime), (I,f)\right]
  + \mu u(I;I^\prime) + \gamma D(f)
\end{equation}
is an supremum function for $L_{\rm cost}(I,f)$.
Therefore, each iterative step becomes
the following minimization problem:
\begin{eqnarray}
  \min_{(I,f)\in R^{M+n_p}} & & \left[ -\sum_{c,p,v}
    \left(\sum_u\left\{z_{c,p,v,u}^{(r)} \log I_u
    - [t_{c,u,v}I_u(\Delta\phi)_p(F_{\rm lt})_{c,p}] \right\}\right.\right.
    \nonumber\\
    & & \left.\left. + w_{c,p,v}^{(r)} \log f_p
    - [f_p d_{0,c,v}(\Delta\phi)_p(F_{\rm lt})_{c,p}]\right)\right.\nonumber\\
    & & \left. + \mu u(I;I^{(r)}) + \gamma D(f)\right]
  \label{eq: min_prob} \\
  & s.t. & \nonumber \\
  -I_u & \le & 0\,\,\,\,(u = 1,\cdots,M), \nonumber\\
  -f_p & \le & 0\,\,\,\,(p = 1,\cdots,n_p) \nonumber.
\end{eqnarray}
If at least one value among $I_u$ and $f_p$ is zero,
the value of equation (\ref{eq: min_prob}) becomes infinity.
So, $I_u \neq 0$ and $f_p \neq 0$.
By using KKT (Karush-Kuhn-Tucker) condition
\citep{Bishop_06, Kanamori+2016}
and introducing Lagrange multipliers
$\mu_u$ and $\tau_p$ ($u = 1,\cdots,M$, $p = 1,\cdots,n_p$),
we get the following relations:
\begin{eqnarray}
  - m_u^{(r)} \frac{1}{I_u}
  + \mu \frac{\partial u(I;I^{(r)})}{\partial I_u}
  + \overline{\epsilon_{u}} - \mu_u & = & 0,
  \label{eq: kkt_main_I} \\
  - n_p^{(r)} \frac{1}{f_p}
  + \gamma \frac{\partial D(f)}{\partial f_p} + \overline{d_p}
  - \tau_p & = & 0,
  \label{eq: kkt_main_f} \\
  \mu_u I_u & = & 0,
  \label{eq: kkt_comp_I}\\
  \tau_p f_p & = & 0,
  \label{eq: kkt_comp_f}
\end{eqnarray}
$I_u \ge 0$, $f_p \ge 0$, $\mu_u \ge 0$, and $\tau_p \ge 0$
$(u = 1, \cdots, M, \,\,\, p = 1, \cdots, n_p)$.
Here,
\begin{eqnarray}
  \overline{\epsilon_{u}} & = &  
  \sum_{c,p,v} t_{c,v,u}(\Delta\phi)_p(F_{\rm lt})_{c,p}, \\
  \overline{d_{p}} & = &
  \sum_{c,v} d_{0,c,v}(\Delta\phi)_p(F_{\rm lt})_{c,p}, \\
  m_u^{(r)} & = & \sum_{c,p,v}z_{c,p,v,u}^{(r)}
  = \sum_{c,p,v} Y_{c,p,v}^{\prime (r)} t_{c,v,u} I_u^{(r)},\\
  n_p^{(r)} & = & \sum_{c,v} w_{c,p,v}^{(r)}
  = \sum_{c,v} Y_{c,p,v}^{\prime (r)} f_p^{(r)} d_{0,c,v}.
\end{eqnarray}
We obtain $\mu_u = 0$($u=1,\cdots,M$) and $\tau_p = 0$($p = 1,\cdots,n_p$)
from equations (\ref{eq: kkt_comp_I},\ref{eq: kkt_comp_f}),
because $I_u \neq 0$ and $f_p \neq 0$.
By the calculation of appendix \ref{sec: appendix uterm},
the derivative of $u(I;I^{(r)})$ can be written by
\begin{equation}
  \frac{\partial u(I;I^{(r)})}{\partial I_u}
  = \alpha_u I_u - \beta_u^{(r)}.
\end{equation}
Finally, from equations (\ref{eq: kkt_main_I}, \ref{eq: kkt_main_f}),
we obtain the updating rule for $I_u$ and $f_p$:
\begin{equation}
  I_u^{(r+1)} = \frac{1}{2\mu \alpha_u}
  \left\{ -(\overline{\epsilon_{u}} - \mu \beta_u^{(r)})
  + \left[(\overline{\epsilon_{u}} - \mu \beta_u^{(r)})^2
    + 4\mu\alpha_u m_u^{(r)} \right]^{1/2} \right\},
\end{equation}
\begin{equation}
  f_p^{(r+1)} = \frac{1}{4\gamma}
  \left\{ -\left(\overline{d_p} - 2\gamma f_{0,p}\right)
  + \left[ \left(\overline{d_p} - 2\gamma f_{0,p}\right)^2
  + 8\gamma n_p^{(r)}\right]^{1/2} \right\}.
\end{equation}
Then, we can solve this minimization problem
by the iterative updating calculation 
for a fixed ($\mu$, $\gamma$).

We show in appendix \ref{sec: appendix proof}
a proof that our algorithm converges
to the unique maximum value from any feasible initial value
($I^{(0)},f^{(0)}$) such that $\log p(I^{(0)},f^{(0)}|Y)$ is a finite value.

\section{Cross-validation of hyper-parameters}

\begin{figure}
  \begin{center}
    \begin{tabular}{ccc}
      & & $\log_{10} \mu$ \\
      & & $-7$ \hspace{2.2cm} $-6$ \hspace{2.2cm} $-5$ \hspace{2.2cm}
      $-4$ \hspace{2.2cm}$-3$\\
      \rotatebox[origin=l]{90}{\hspace{3.6cm} $\log_{10}\gamma$} &
      \rotatebox[origin=l]{90}{\hspace{0.8cm} $-5$ \hspace{2.2cm}
        $-7$ \hspace{2.2cm} $-9$}
      & \includegraphics[bb=0 0 721 425, width=15cm]{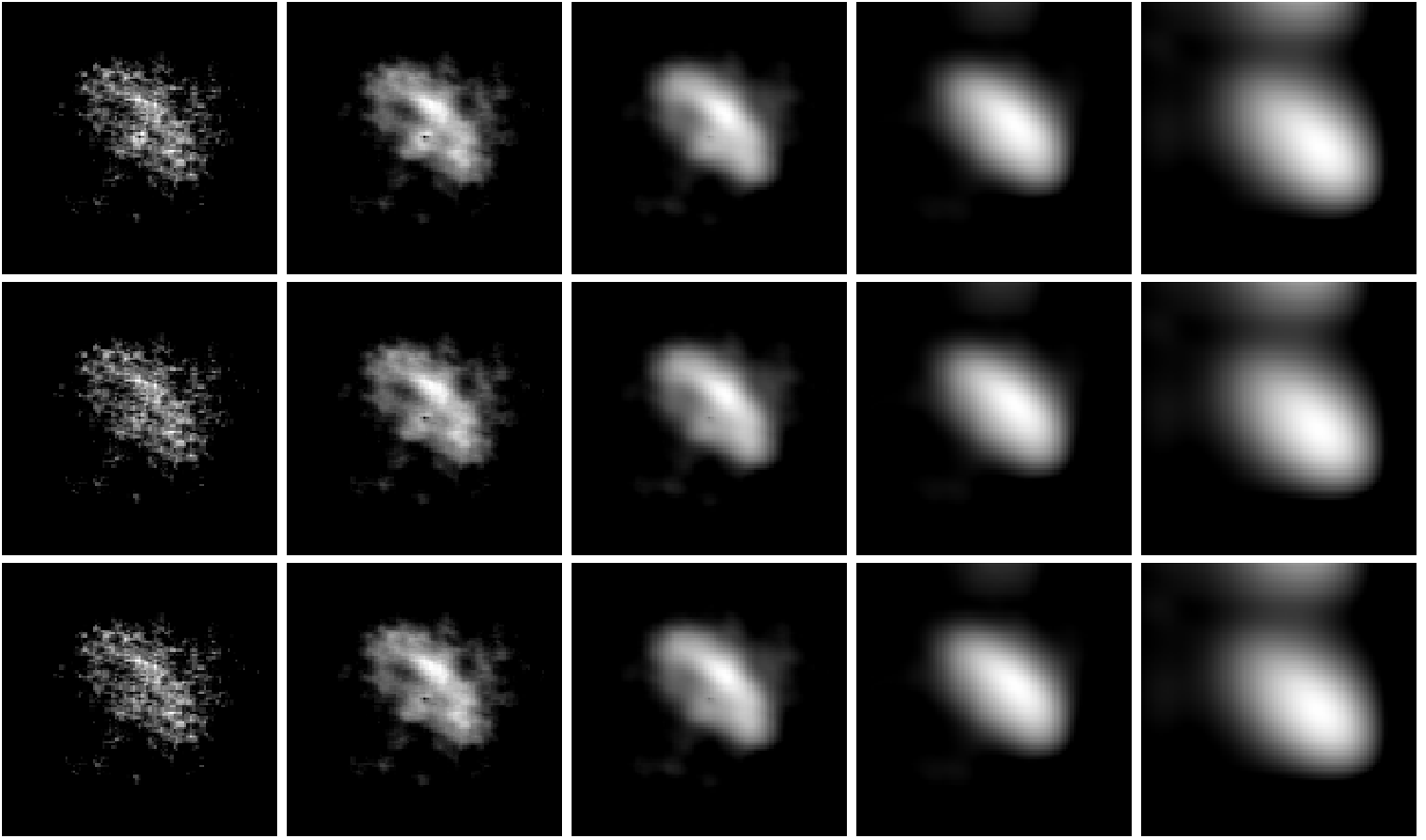}
    \end{tabular}
 \end{center}
  \caption{Deconvolved images of the Crab Nebula made from
    the data of {\it Hitomi} HXI-1 in $3.6 - 15$keV band
    without the pulsar component, by changing the hyper-parameters
    ($\log_{10}\mu$, $\log_{10}\gamma$). These values are shown 
    above and to the left of the figure, respectively.
    The best image determined by the cross-validation is
    the one at ($-5$, $-9$).}
\label{fig: mu_gamma}
\end{figure}

Since the deconvolved image depends
on the hyper-parameters ($\mu$, $\gamma$; see figure
\ref{fig: mu_gamma} for example),
we then need determine the parameters by the cross-validation as follows.
We apply five-fold cross-validation, in which
the image data are randomly divided into five sub-data
with equivalent photon counts and
five pairs of training and validation data are made,
where the training one is made up of four sub-data and
the validation one is made up of the remaining sub-data.
Then, for each pair, the nebula image and pulsar fluxes
are obtained by the deconvolution using one of training data
and the quality is evaluated against the corresponding validation data.
We use the root mean squared error (RMSE)
between the image on the detector
convolved for the deconvolved image
made from one of training data and the corresponding validation data.
Here, the RMSE is calculated to be
\begin{equation}
  {\rm RMSE} = \left\{ \frac{ \sum_{p=1}^{n_p}\sum_{c=1}^{n_c}
    \left[({\rm RMSE})_{c,p}\right]^2 }
  {n_c n_p} \right\}^{1/2},
\end{equation}
where
\begin{eqnarray}
  ({\rm RMSE})_{c,p} & = & \left( \frac{1}{V}
    \sum_{v = 1}^V \left\{
    \frac{1}{N_{\rm fold} - 1} \left[ \sum_u 
      t_{c,v,u}(I_u^{(\rm deconv)} + f_p^{(\rm deconv)} I_{0,u})
      + b_{c,v} \right]
    (\Delta \phi)_p (F_{\rm lt})_{c,p} \right.\right.\nonumber \\ 
    & & \left.\left. - Y_{c,v}^{(\rm val)}
    \right\}^2
    \right)^{1/2},
\end{eqnarray}
and $N_{\rm fold} = 5$.
$I_u^{(\rm deconv)}$, $f_p^{(\rm deconv)}$,
and $Y_{c,v}^{(\rm val)}$ are
the deconvolved nebula image, the pulsar flux,
and the validation data, respectively.
The average and variance of the RMSE are
calculated over the five training-validation data pairs.
The evaluation is repeated for each hyper-parameter pair ($\mu$, $\gamma$).

The $\gamma$ parameter mainly affect the pulsar flux
and has little effect on the nebula image (figure \ref{fig: mu_gamma}).
We then adapt the parameter with the smallest RMSE
for this parameter.
For the $\mu$ parameter instead of adapting
at the smallest RMSE, we adapt the one-standard error rule
to avoid the over-training for the training data [for example,
\citet{Hastie_Tibshirani_Wainwright_2015}].
In this rule, we adapt the largest $\mu$ (the smoothest) parameter,
allowed within the fluctuation of the RMSE.
In other words, we adapt the $\mu$ value at which
the RMSE is the same as the value obtained
by the addition of the minimum RMSE and the one-standard error
of RMSE at the $\mu$ with the minimum RMSE.
We search the parameter among 121 pairs of ($\mu$, $\gamma$),
where $\log_{10} \mu$ and $\log_{10} \gamma$
are varied from $-10$ to $0$ by a step of $1$.

We implement the above algorithm with C++, utilizing the BLAS library
\footnote{https://www.netlib.org/blas/}.
For each fixed hyper-parameter ($\mu$, $\gamma$),
we start the deconvolution calculation
from an initial value ($I^{(0)}, f^{(0)}$)
of a flat image and a constant pulse profile,
and the updating calculation stops after 10000 iterations or
when the Hellinger distance between the current and previous images
becomes less than $10^{-7}$.
The processing speed is measured by using a computer
equipped with CPUs of AMD EPYC 7543.
It has 32 cores and the clock is 2.8 GHz.
The nebula image with $100 \times 100$ pixels and
10 pulse-phase fluxes are deconvolved 
by an average of 755~s with one standard deviation of 244~s
for each hyper-parameter,
when using 1 core.
For the fast computation, we use CUDA
\footnote{https://docs.nvidia.com/cuda/}
working on GPU (NVIDIA RTX A4000),
then mark about 20-fold speed-up.

\section{Results}

\begin{figure}
  \begin{center}
    \begin{tabular}{cc}
      & $3.6 - 15$keV (HXI-1) \hspace{0.5cm} $15-30$ keV (HXI-1,2)
      \hspace{0.5cm} $30 - 70$keV (HXI-1,2)\\
      \rotatebox[origin=l]{90}
                {\hspace{2.5cm} Deconvolved \hspace{4.0cm} Observed}&
      \includegraphics[bb=0 0 511 425, width=15cm]
                      {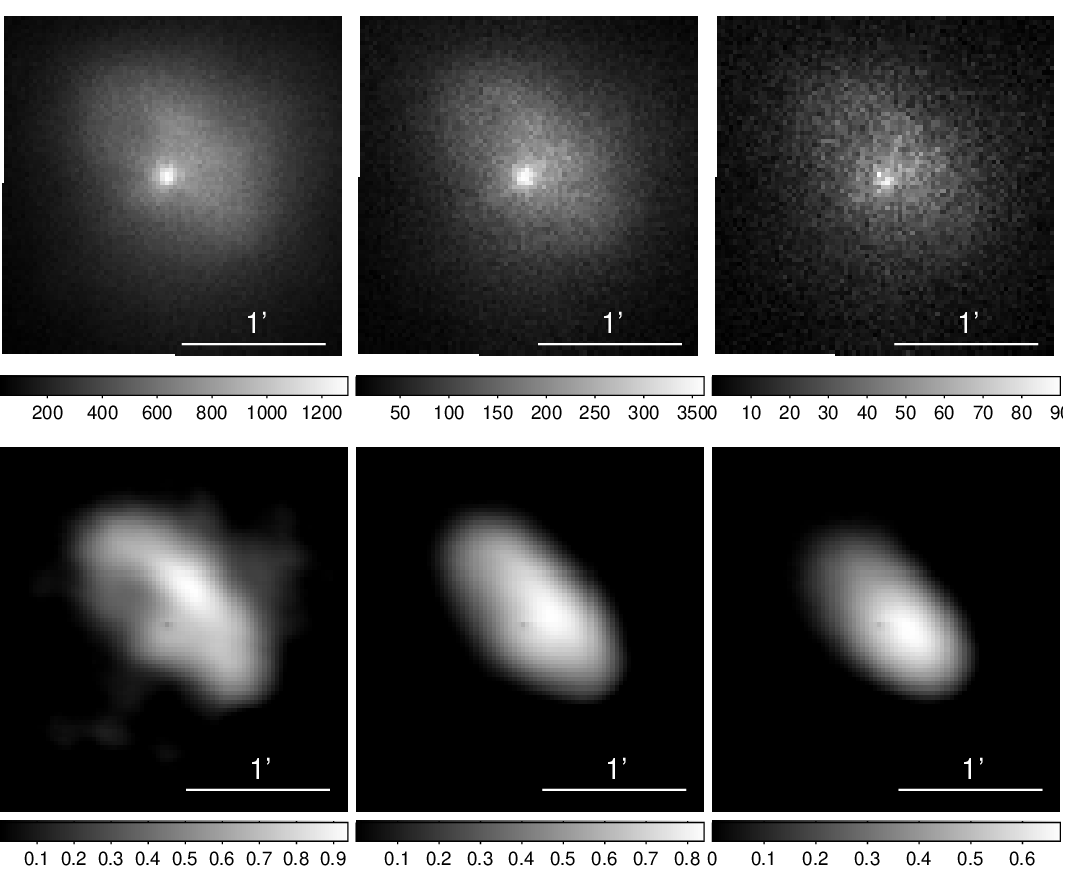}
    \end{tabular}
  \end{center}
 \caption{The observed and deconvolved images of the Crab Nebula
   obtained by {\it Hitomi} HXI in three energy bands.
   The upper and lower panels show the observed images taken by HXI
   and the deconvolved images without the pulsar component, respectively.
   The left panels show $3.6 - 15$ keV band images using HXT-1 data.
   The center and right panels show $15 - 30$ keV and
   $30 - 70$ keV band images, respectively, where
   both HXT-1 and HXT-2 data are used.
   The deconvolved images are the results with the hyper-parameters
   of $(\log_{10}\mu, \log_{10}\gamma) = (-5,-9)$, $(-4,-8)$ and $(-3,-7)$,
   from left to right, in which $\mu$ parameters are determined
   by the one-standard error rule.
   The size of panel is $80 \times 80$ pixels,
   and the horizontal line segment shows
   the size of 1 arcmin in each panel.
 }
\label{fig: reconst}
\end{figure}

\begin{figure}
  \begin{center}
    \includegraphics[bb=0 0 841 425, width=15cm]
                    {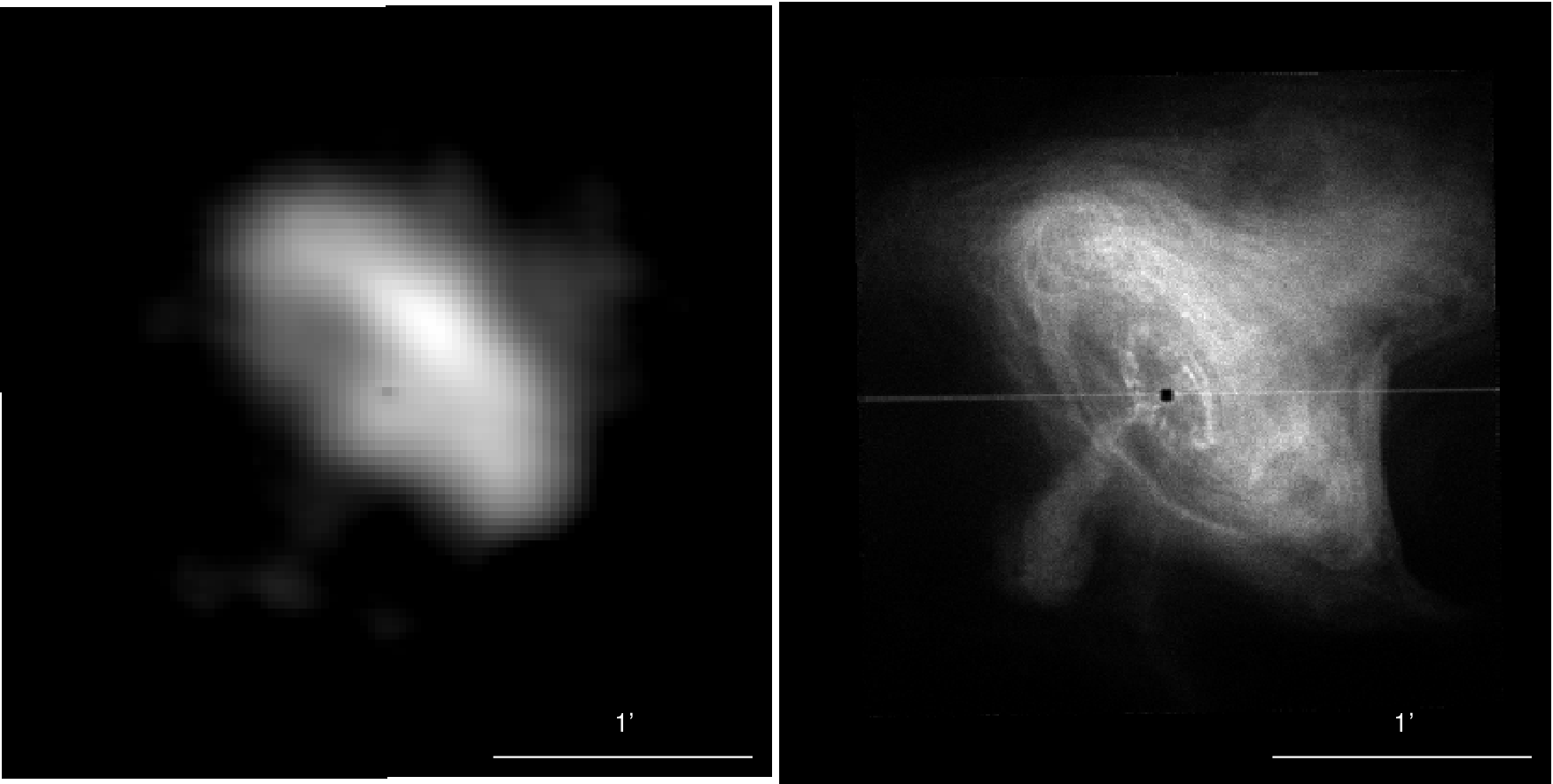}
  \end{center}
  \caption{Comparison between
    the deconvolved {\it Hitomi} HXI-1 $3.6 - 15$ keV band
    image (left) and {\it Chandra} ACIS $0.5 - 7.0$ keV band image (right).
    The horizontal line segment shows the size of 1 arcmin in each panel.
 }
\label{fig: hitomi_chandra}
\end{figure}

We apply the above deconvolution method for the {\it Hitomi} observation data
for three energy bands of $3.6-15$, $15-30$ and $30-70$ keV.
The observed {\it Hitomi} HXI images in these energy bands
are shown in the upper panels of figure \ref{fig: reconst}.
By the cross-validation, we determine the best hyper-parameter 
and apply the deconvolution with fixing the hyper-parameters.
The resultant deconvolved images are shown
in the lower panels of figure \ref{fig: reconst}.

As a demonstration of our method,
figure \ref{fig: hitomi_chandra} shows a comparison
between the deconvolved image in $3.6 - 15$ keV band
and {\it Chandra} ACIS $0.5 - 7.0$ keV band image \citep{Mori+2004}.
Our {\it Hitomi} HXI image looks similar
with the {\it Chandra} image. The torus, south-east jet
and north-west extended emission are also seen in our deconvolved image.

\section{Evaluation of systematic uncertainty
  of the response matrix}
\label{sec: uncertainty}

The response matrix used is expected to contain the following
three kinds of uncertainties.
We evaluate the effects for the deconvolved image.

\subsection{Uncertainty of livetime fraction}

Table~\ref{tab: hxt_deadtime} shows the livetime fraction for each phase,
calculated by the method in the appendix of \citet{Matsumoto+2018}.
It confirms that the livetime fraction at each phase is
about three-fourths, and that the ON phase image has
less livetime fraction than the OFF, because the phase of
higher count rate has the lower livetime fraction.
\citet{Matsumoto+2018} also reported that the uncertainty on
the deadtime fraction ($F_{\rm dt}$)
at the 1$\sigma$ confidence level is $\Delta F_{\rm dt} / F_{\rm dt} = 2$\%.
So, the uncertainty of the livetime fraction ($F_{\rm lt}$) is
$\Delta F_{\rm lt} / F_{\rm lt}
= (F_{\rm dt} / F_{\rm lt}) \times (\Delta F_{\rm dt} / F_{\rm dt})
\sim (0.25/0.75) \times 2 \% = 0.7\%$.

\begin{table}
  \caption{Livetime fraction of each pulse phase ID}
  \begin{center}
    \begin{tabular}{lcccc} \hline
      Phase ID & Phase & Phase duration
      & \multicolumn{2}{c}{Livetime fraction}   \\
          &              &      & HXI-1 & HXI-2 \\ \hline
      All & 0-1          & 1    & 0.766 & 0.752 \\
      ON2 & 0.854-1.104  & 0.25 & 0.750 & 0.735 \\
      OFF1 & 0.104-0.504 & 0.40 & 0.783 & 0.770 \\
      ON1 & 0.504-0.704  & 0.20 & 0.747 & 0.731 \\
      OFF2 & 0.704-0.854 & 0.15 & 0.773 & 0.758 \\ \hline
    \end{tabular}
  \end{center}
  \label{tab: hxt_deadtime}
\end{table}

\begin{figure}
  \begin{center}
    \begin{tabular}{ccc}
      \includegraphics[width=5.0cm]{
        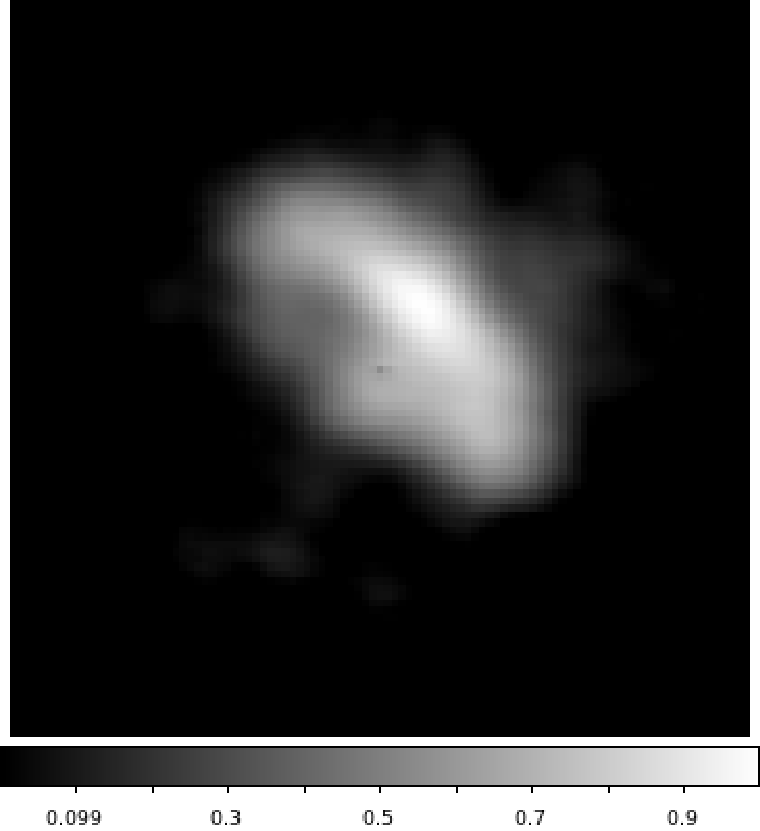} &
      \includegraphics[width=5.0cm]{
        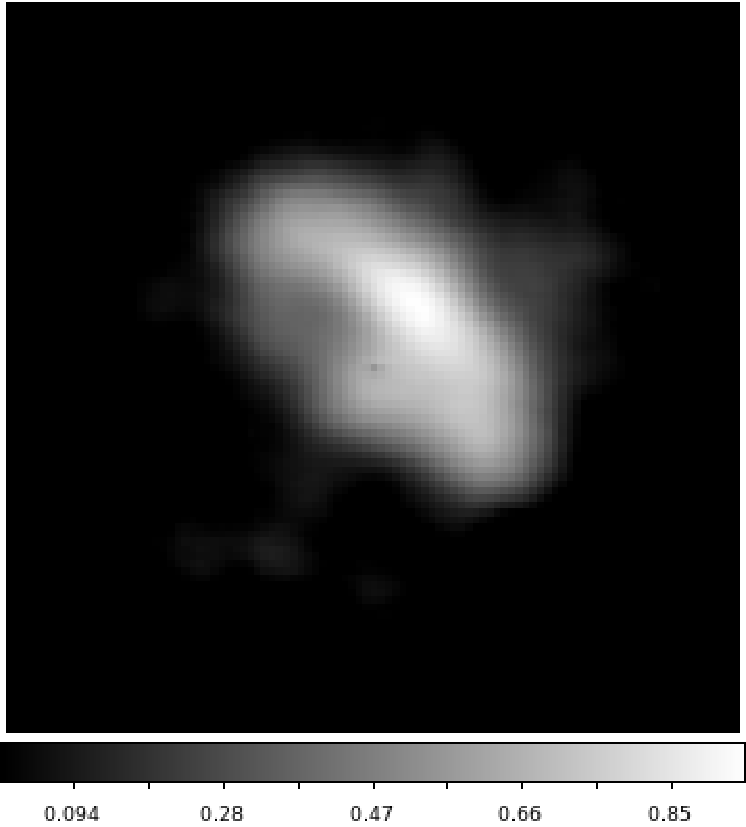} &
      \includegraphics[width=5.0cm]{
        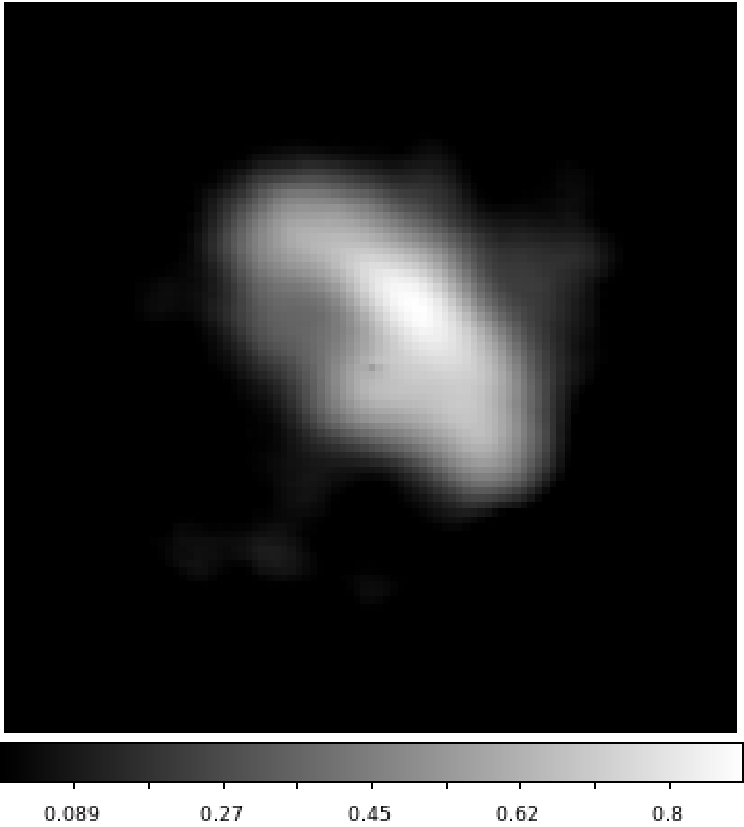}
    \end{tabular}
  \end{center}
  \caption{The deconvolved images of the Crab Nebula
    in the 3.6--15 keV band, using three different response matrices,
    made by assuming different livetime fraction for the OFF1 image.
    Here, the cases of 2\% decrease, nominal and 2\% increase
    of the livetime fraction are shown in left, center and right panels,
    respectively.
  }
  \label{fig: livetimesub}
\end{figure}

Since the PSF is made by the image subtraction of
ON minus OFF1 pulse phase,
the uncertainty of the livetime fraction causes
the uncertainty for the shape of the subtracted image.
In order to evaluate the effect of the uncertainty
for the deconvolved image conservatively,
we increase/decrease the livetime of OFF1 image by $2$\%,
corresponding to $3 \times \sim 0.7$\%,
make the PSF by the ON-OFF1 operation, and
deconvolve the Crab Nebula image in the 3.6-15 keV band.
Figure~\ref{fig: livetimesub} shows the deconvolved images made
by using the response matrices in three different livetime cases.
There are no significant difference among these images.
We then conclude that the livetime uncertainty
does not affect the deconvolved images.

\subsection{Uncertainty of optical axis location}

The HXT shows a sharp vignetting effect as reported in \citet{Awaki+2014}.
They showed that the effective area decreases by 10--20\%
(dependent on energy) at 1 arcmin off-axis from the optical axis.
Therefore, the uncertainty of the optical axis location
can affect the shape of the deconvolved image.
By the energy spectral analysis
of \citet{Matsumoto+2018} and \citet{Hagino+2018},
the location of the Crab pulsar is confirmed to be
within 0.5 arcmin from the optical axis.
We thus evaluate the difference of the deconvolved images
made by changing the exposure maps corresponding to
the different optical axis locations,
within the conservative uncertainty of 1 arcmin
from the pulsar position.

\begin{figure}
  \begin{center}
    \begin{tabular}{ccc}
      & & Energy band [keV] \\
      & & $3.6$--$15$ \hspace{2.8cm} $15$--$30$ \hspace{3.2cm}
      $30$--$70$ \\
      \rotatebox[origin=l]{90}{\hspace{3.6cm}
        Position of an assumed optical axis} &
      \rotatebox[origin=l]{90}{\hspace{2.3cm}
        bottom right \hspace{2.5cm}
        center \hspace{2.5cm} upper left}
      &  \includegraphics[bb=0 0 550 530, width=15cm]{
        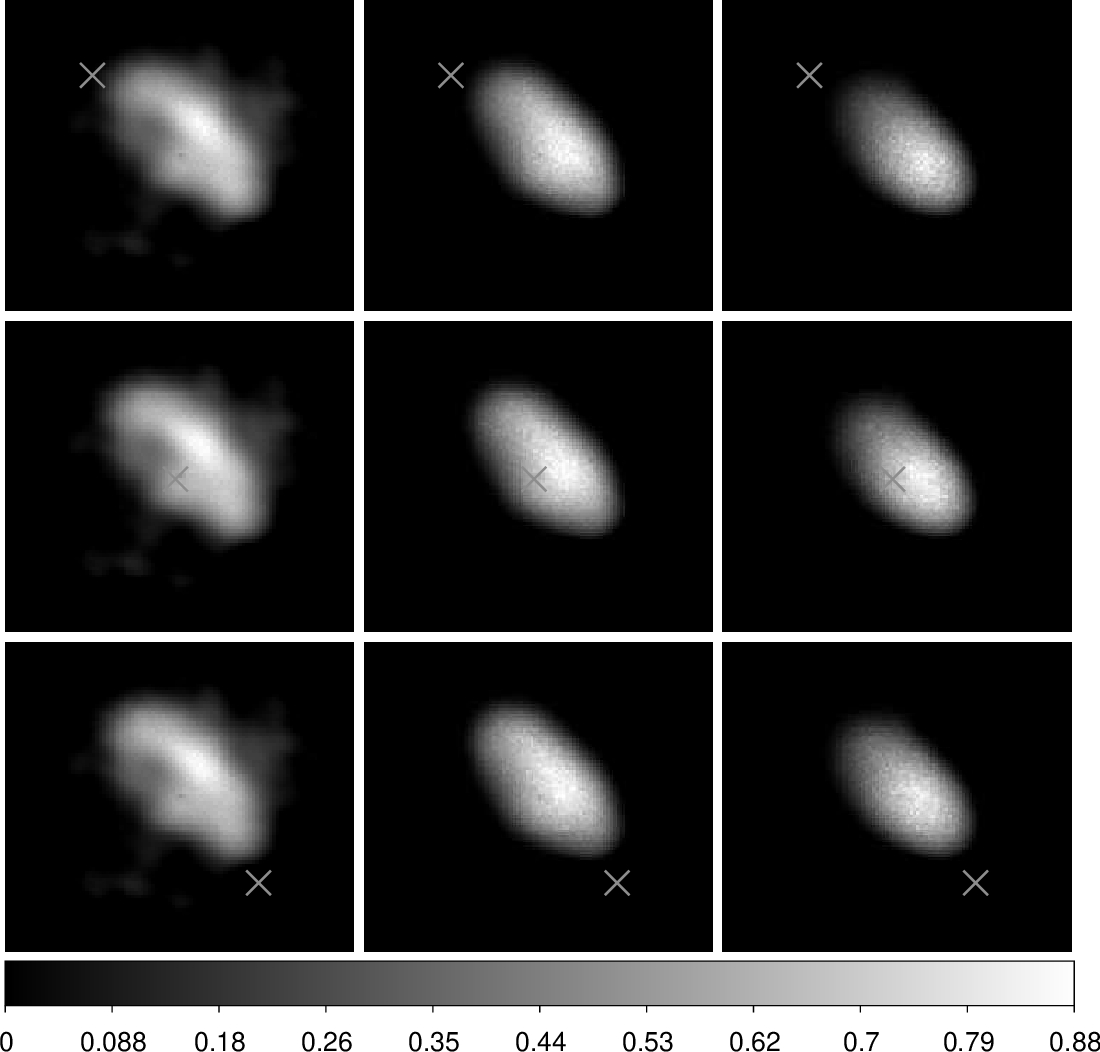}
    \end{tabular}
  \end{center}
  \caption{
    Exposure-corrected deconvolved images in the 3.6–15 (left column),
    15–30 (center column) and 30–70 (right column) keV bands
    for three different optical-axis locations.
    The cross mark is the optical-axis location we assumed
    to make the exposure map. The middle row images are deconvolved
    by assuming the optical axis is located at the pulsar position.
    In the upper and lower panels, the optical axes are located 1 arcmin
    to north-east and south-west directions from the pulsar position,
    respectively.
  }
  \label{fig: vigeff}
\end{figure}

We make the exposure maps assuming three different optical
axis locations using the ray-tracing code {\it xrrtraytrace}
(\cite{Angelini+2016}) for each energy band.
The deconvolved images corrected for these exposure maps
are shown in figure~\ref{fig: vigeff},
where the optical axis is located at 1 arcmin shifted
to north-east direction (top row),
at the pulsar position (middle row),
and 1 arcmin shifted to south-west direction (bottom row).
In figure~\ref{fig: vigeff}, no remarkable difference
are seen among different optical axes and these energy bands.
Thus, we conclude that the uncertainty of the optical axis location
does not affect the deconvolved images.

\subsection{Uncertainty of the off-axis PSF}

The off-axis effect is known to not only reduce the effective area
but also change the PSF \citep{Awaki+2014}.
Since the drop of the effective area at 1 arcmin
off-axis location is only 10-20\%,
it contributes small change of the PSF.
It is also known that the conical approximation adapted for the HXT
shows negligible off-axis dependence of the HPD
of the PSF \citep{Petre+1985}.
Thus, we decide to adapt the Crab pulsar PSF
to all in-coming directions in this paper.

\section{Image analysis}

\begin{figure}
  \begin{center}
    \begin{tabular}{ccc}
      \includegraphics[width=5cm]{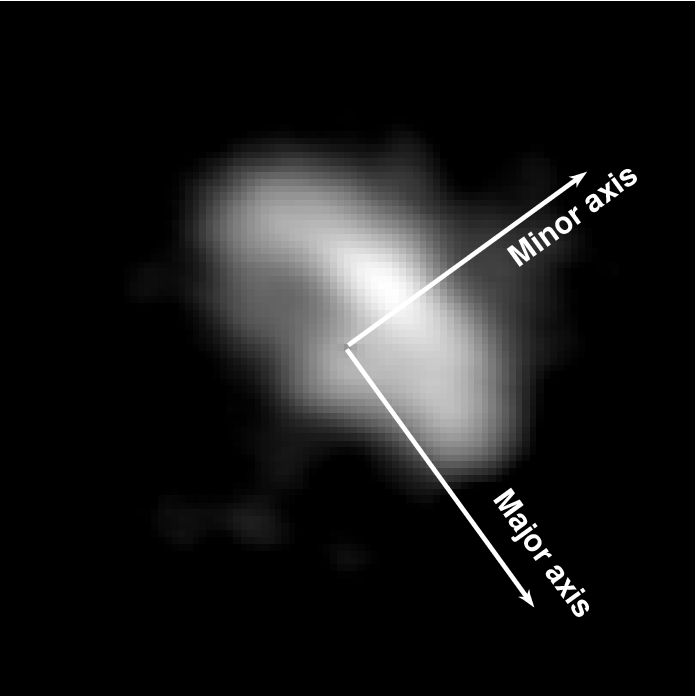} &
      \includegraphics[width=5.5cm]{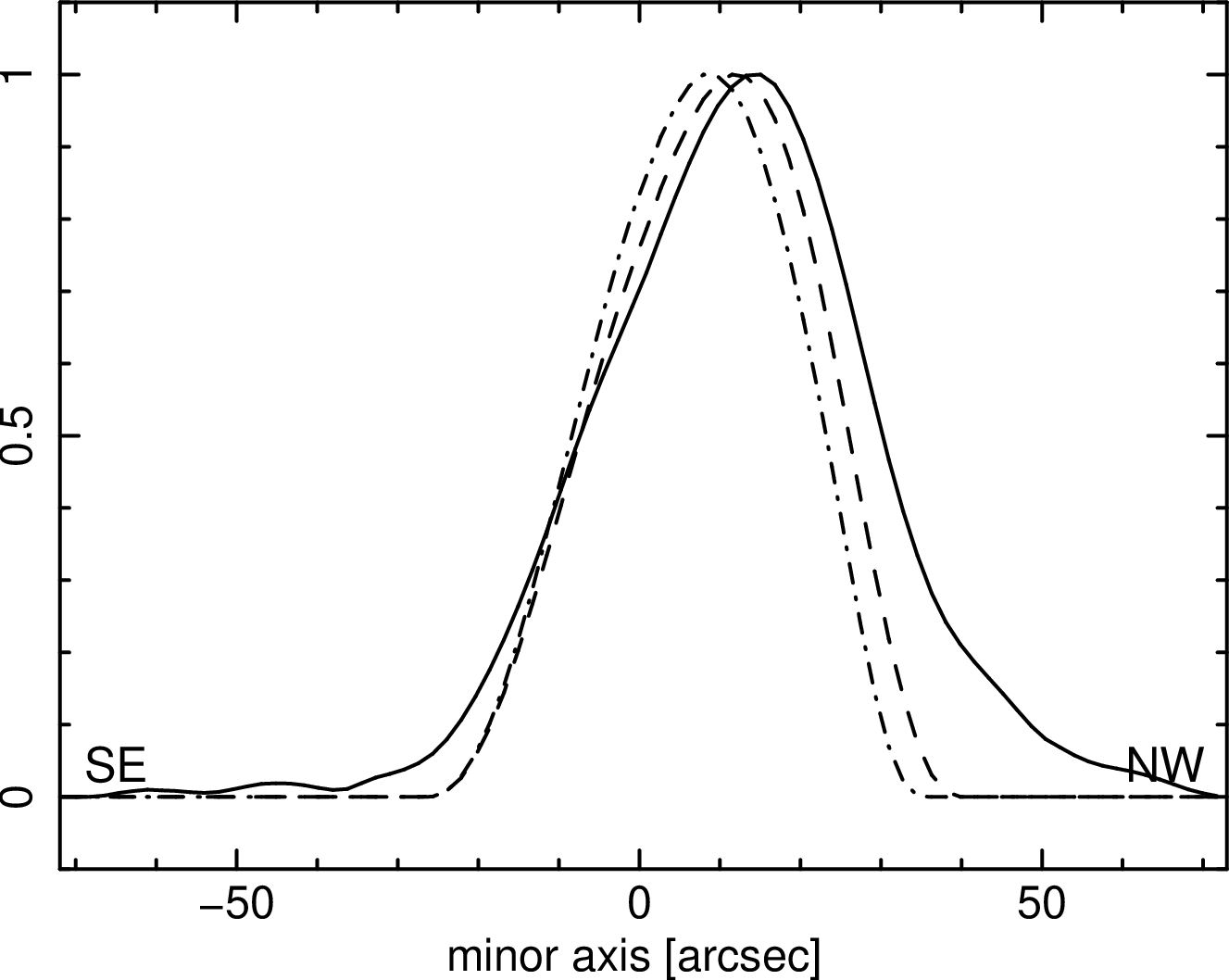} &
      \includegraphics[width=5.5cm]{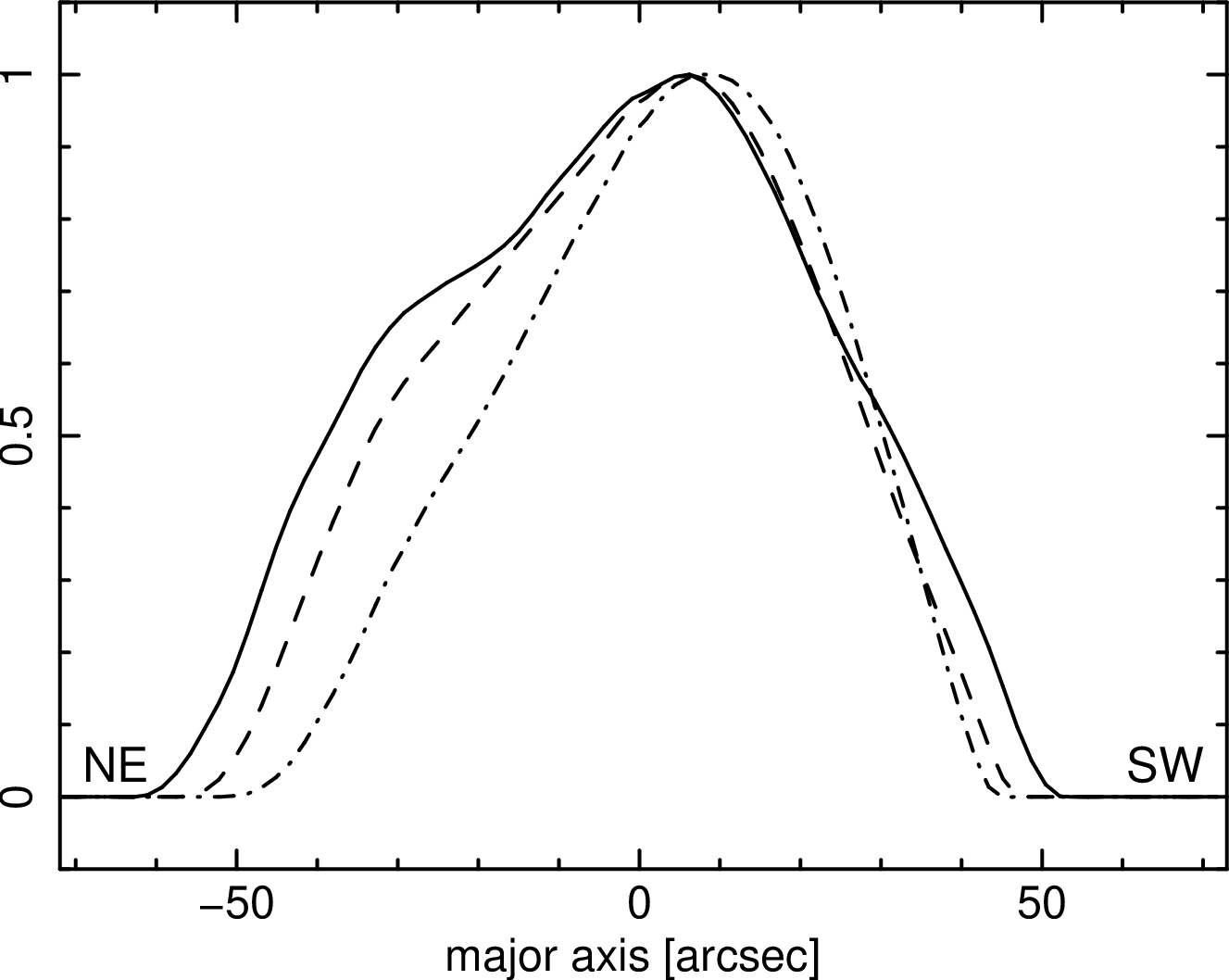} \\
    \end{tabular}
  \end{center}
  \caption{
    Projections of the deconvolved images along the minor and major axes
    (center and right, respectively),
    whose axes are shown in the left image (3.6-15 keV band).
    The origins are at the pulsar position.
    The solid, dash and dash-dot lines correspond
    to the energy bands of 3.6--15, 15--30 and 30--70 keV,
    respectively. The peaks are normalized by unity.
  }
  \label{fig: projections}
\end{figure}

As shown in figure \ref{fig: reconst},
the nebula size is obviously smaller in higher energy bands.
In order to measure the nebula size,
we project the image along the minor and major axes
as shown in figure~\ref{fig: projections} (left),
which are rotated clockwise by 54 degrees
from the north and west directions, respectively.
Here, the minor axis is determined to align the pulsar spin axis
obtained by the {\it Chandra} image.
The projected images in these axes
are shown in figure \ref{fig: projections} (center and right).
The nebula sizes in these axes are listed in table \ref{tab: projections}.
They clarify such the trend that the nebula
is smaller in higher energy bands.
Here, we need mention that
the higher energy bands have smaller number of photons,
then the smoothing parameter ($\mu$) tends to be large,
and so the resulting nebula image tends to be large.
Nevertheless, the harder band images show smaller nebula size.
Thus, this trend is strongly supported.
It was also pointed out by the {\it NuSTAR}
using the OFF1 phase data \citep{Madsen+2015}.

\begin{table}
  \caption{Nebula size evaluated by FWHM
    in three energy bands}
  \begin{center}
    \begin{tabular}{ccc} \hline
      Energy band & Minor axis & Major axis\\
      keV--keV & Pixels (arcsec) & Pixels (arcsec) \\ \hline
      3.6--15 & 22(39) & 41(72) \\
      15--30 & 20(35) &36(64) \\
      30--70 &19(34) &30(53) \\ \hline
    \end{tabular}
  \end{center}
  \label{tab: projections}
\end{table}

Along the minor axis, the north-west (NW) direction is brighter than
the south-east (SE) direction in all energy bands.
It is explained by the relativistic viewing effect of
the high energy electron/positron at the torus
(\cite{Mori+2004} and reference therein).
Along the major axis, the north-east (NE) direction is darker
than the south-west (SW) direction in higher energy band.
This trend is reported for the first time.

\section{Discussion and future perspective}

Our deconvolution image of the {\it Hitomi} HXI
in the 3.6--15 keV band succeeds to clarify
the torus-like structure including its inner boundary.
The structure does not appear in the deconvolved image
of the {\it NuSTAR} data using the Richardson-Lucy method
\citep{Madsen+2015}.
Both {\it NuSTAR} and {\it Hitomi} have
nearly equal angular resolution in HPD.
So, the difference is thought to be caused by two advantages in our work.
The first one is our ingenious method
that introduces two components for the nebula and pulsar
with regularization for smoothness and flux, respectively,
and the multi-pulse-phase simultaneous deconvolution.
The second one is the sharp core of the PSF of the {\it Hitomi} HXT,
which is smaller by a factor of two than that of {\it NuSTAR}.
It simply improves the angular resolution of the deconvolved image.

We can not clearly identify the torus-like structure
in the hard X-ray image using the same method above 15 keV.
We however confirm the {\it NuSTAR}'s findings that
the size of the Crab Nebula decreases in higher energy bands.
\citet{Madsen+2015} pointed out the discrepancy
of the averaged photon index of the nebula:
$\sim1.9$ \citep{Mori+2004} and $\sim2.140\pm0.001$ above 100~keV
\citep{Pravdo_Angelini_Harding_1997}.
They also pointed out that the discrepancy is due to the spectral cut-off
of the outer torus around the 10–100~keV band.
Using the {\it Hitomi} data, we found that
the north-east becomes darker in the higher energy bands,
indicating the spectra of the north-east torus is more rapidly cut off
than that of the south-west.
We expect our result will promote
the theoretical works of the Crab Nebula.

Our deconvolution algorithm can be applicable for
any telescope images of faint diffuse objects
containing a bright point source,
and effectively works especially for the
case that the apparent size of the diffuse objects is
comparable with that of the PSF of the telescope.
We also mention that the algorithm can be extended for objects
with multiple point sources.
Another extension would be possible to introduce
smoothness constraint for the energy direction.
The calculation speed would be improved by more tuning in the CUDA coding.
It is also improved by introducing some acceleration methods
of the MM algorithm. Indeed, we observed the decrease of
the number of iterations by the acceleration methods
\citep{Varadhan_Roland_2008, Zhou_Alexander_Lange_2011}.
For the hyper-parameter tuning,
the Bayesian optimization is a promising method \citep{Garnett_2023}.
We make our source code open for public use
\footnote{The github address: https://github.com/moriiism/srt/}.

%



\begin{ack}
  This research is supported by JSPS Grants-in-Aid for
  Scientific Research (KAKENHI) Grants No.~22H01277 and 21H01095.
\end{ack}

\appendix 

\section{Detailed calculation}
\label{sec: appendix uterm}

The surrogate function for $V(I)$ function is
\begin{eqnarray}
  u(I;I^\prime) & = & \frac{1}{2} \sum_{(r,s)\in N}
  \{[2I_r - (I_r^\prime + I_s^\prime)]^2
  + [2I_s - (I_r^\prime + I_s^\prime)]^2 \} \nonumber\\
  & = & \frac{1}{2} \left( \sum_{i=1}^{m-1}\sum_{j=1}^{n-1}
  \{ [2I_{i,j}-(I_{i,j}^\prime + I_{i+1,j}^\prime)]^2
  + [2I_{i+1,j}-(I_{i,j}^\prime + I_{i+1,j}^\prime)]^2
  \right. \nonumber\\
  & & + [2I_{i,j}-(I_{i,j}^\prime + I_{i,j+1}^\prime)]^2
  + [2I_{i,j+1}-(I_{i,j}^\prime + I_{i,j+1}^\prime)]^2 \} \nonumber\\
  & & + \sum_{i=1}^{m-1} \{
  [2I_{i,n}-(I_{i,n}^\prime + I_{i+1,n}^\prime)]^2
  + [2I_{i+1,n}-(I_{i,n}^\prime + I_{i+1,n}^\prime)]^2 \} \nonumber\\
  & & \left. + \sum_{j=1}^{n-1} \{
  [2I_{m,j}-(I_{m,j}^\prime + I_{m,j+1}^\prime)]^2
  + [2I_{m,j+1}-(I_{m,j}^\prime + I_{m,j+1}^\prime)]^2 \}
  \right)
\end{eqnarray}
and the derivative of $u(I;I^\prime)$ is
\begin{eqnarray}
  \frac{\partial u(I;I^\prime)}{\partial I_{kl}} & = &
  \frac{1}{2}\left(
  \sum_{i=1}^{m-1}\sum_{j=1}^{n-1}\{
  2[2I_{i,j} - (I_{i,j}^\prime + I_{i+1,j}^\prime)] 2\delta_{i,k}\delta_{j,l}
  + 2[2I_{i+1,j} - (I_{i,j}^\prime + I_{i+1,j}^\prime)] 2\delta_{i+1,k}\delta_{j,l}
  \right. \nonumber \\
  & &+ 2[2I_{i,j} - (I_{i,j}^\prime + I_{i,j+1}^\prime)] 2\delta_{i,k}\delta_{j,l}
  + 2[2I_{i,j+1} - (I_{i,j}^\prime + I_{i,j+1}^\prime)] 2\delta_{i,k}\delta_{j+1,l}
  \} \nonumber\\
  & & + \sum_{i=1}^{m-1}\{
  2[2I_{i,n} - (I_{i,n}^\prime + I_{i+1,n}^\prime)] 2\delta_{i,k}\delta_{n,l}
  + 2[2I_{i+1,n} - (I_{i,n}^\prime + I_{i+1,n}^\prime)] 2\delta_{i+1,k}\delta_{n,l}
  \} \nonumber\\
  & & \left. + \sum_{j=1}^{n-1}\{
  2[2I_{m,j} - (I_{m,j}^\prime + I_{m,j+1}^\prime)] 2\delta_{m,k}\delta_{j,l}
  + 2[2I_{m,j+1} - (I_{m,j}^\prime + I_{m,j+1}^\prime)] 2\delta_{m,k}\delta_{j+1,l}
  \}\right) \nonumber \\
  & = & ({\rm term 1}) + ({\rm term 2}) + ({\rm term 3}) + ({\rm term 4})
  \nonumber \\
  & & + ({\rm term 5}) + ({\rm term 6}) + ({\rm term 7}) + ({\rm term 8}).
\end{eqnarray}
Here,
\begin{eqnarray}
  ({\rm term 1}) & = & 
  \left\{
  \begin{array}{cc}
    2 [2I_{k,l} - (I_{k,l}^\prime + I_{k+1,l}^\prime)] &
    (1 \le k \le m-1, 1 \le l \le n-1) \\
    0 & ({\rm otherwise})
  \end{array}
  \right. , \\
  ({\rm term 2}) & = &
  \left\{
  \begin{array}{cc}
    2 [2I_{k,l} - (I_{k-1,l}^\prime + I_{k,l}^\prime)] &
    (2 \le k \le m, 1 \le l \le n-1) \\
    0 & ({\rm otherwise})
  \end{array}
  \right. , \\
  ({\rm term 3}) & = &
  \left\{
  \begin{array}{cc}
    2 [2I_{k,l} - (I_{k,l}^\prime + I_{k,l+1}^\prime)] &
    (1 \le k \le m-1, 1 \le l \le n-1) \\
    0 & ({\rm otherwise})
  \end{array}
  \right. , \\
  ({\rm term 4}) & = &
  \left\{
  \begin{array}{cc}
    2 [2I_{k,l} - (I_{k,l-1}^\prime + I_{k,l}^\prime)] &
    (1 \le k \le m-1, 2 \le l \le n) \\
    0 & ({\rm otherwise})
  \end{array}
  \right. , \\
  ({\rm term 5}) & = &
  \left\{
  \begin{array}{cc}
    2 [2I_{k,n} - (I_{k,n}^\prime + I_{k+1,n}^\prime)] &
    (1 \le k \le m-1, l = n) \\
    0 & ({\rm otherwise})
  \end{array}
  \right. , \\
  ({\rm term 6}) & = &
  \left\{
  \begin{array}{cc}
    2 [2I_{k,n} - (I_{k-1,n}^\prime + I_{k,n}^\prime)] &
    (2 \le k \le m, l = n) \\
    0 & ({\rm otherwise})
  \end{array}
  \right. , \\
  ({\rm term 7}) & = &
  \left\{
  \begin{array}{cc}
    2 [2I_{m,l} - (I_{m,l}^\prime + I_{m,l+1}^\prime)] &
    (k = m, 1 \le l \le n-1) \\
    0 & ({\rm otherwise})
  \end{array}
  \right. , \\
  ({\rm term 8}) & = &
  \left\{
  \begin{array}{cc}
    2 [2I_{m,l} - (I_{m,l-1}^\prime + I_{m,l}^\prime)] &
    (k = m, 2 \le l \le n) \\
    0 & ({\rm otherwise})
  \end{array}
  \right. .
\end{eqnarray}


\section{Proof of convergence of our algorithm}
\label{sec: appendix proof}

Here, we proof convergence of our algorithm,
partly following the proof shown in \citet{Kanamori+2016}.
The cost function $L_{\rm cost}(x)$ is defined at a non-negative orthant,
where $x = (I, f) \in \{x \in R^{M + n_p}|
x_i \ge 0 \,\, (i = 1, 2, \cdots, M + n_p)\}$.
It is a continuous proper convex function, and then closed
[\citet{Rockafellar_1970}, section 7].
It becomes infinity when $|x| \rightarrow \infty$, then
it has no directions of recession [\citet{Rockafellar_1970}, section 8].
Because of \citet{Rockafellar_1970}, Theorem 27.1 (d),
the minimum set of $L_{\rm cost}$ is a non-empty bounded closed convex set.
All the level set
${\rm lev}_\alpha L_{\rm cost}
= \{x \in R^{M + n_p}| L_{\rm cost}(x) \le \alpha \}$
($\alpha \in {\rm R}$) is
a bounded closed convex set [\citet{Rockafellar_1970}, Theorem 27.1 (f)
and Theorem 8.4], and  then a compact set.
Since $\mu V(I) + \gamma D(f)$ is strictly convex
and $-\log p(Y|I,f)$ is convex,
$L_{\rm cost}(x)$ is also a strictly convex function in any
level set [\citet{Rockafellar_1970}, section 26].
So, the minimum set of $L_{\rm cost}$ cannot contain
more than one point [\citet{Rockafellar_1970}, Section 27],
then the minimum set of $L_{\rm cost}$ is made up of a unique point.

The MM algorithm produce a sequence $(x^{(r)})_{r = 1,2,\cdots}$ 
from any feasible initial value $x^{(0)}$
such that $L_{\rm cost}(x^{(0)})$ is finite.
Since the level set 
${\rm lev}_{\rm init} =
\{x \in R^{M + n_p} | L_{\rm cost}(x) \le L_{\rm cost}(x^{(0)})\}$
is a compact set, there exists a sub-sequence
$(x^{(r_k)})_{k = 1,2,\cdots}$
which converges to a value within the level set:
$\lim_{k \rightarrow \infty} x^{(r_k)} = x^* \in {\rm lev}_{\rm init}$.
Since the MM algorithm produces a monotonically decreasing sequence:
$L_{\rm cost}(x^{(r)}) \ge L_{\rm cost}(x^{(s)})\,\,(r < s)$,
$u_{\rm cost}(x^{(r_{k+1})}; x^{(r_{k+1})}) = L_{\rm cost}(x^{(r_{k+1})})
\le L_{\rm cost}(x^{(r_k + 1)})
\le u_{\rm cost}(x^{(r_k + 1)}; x^{(r_k)})
\le u_{\rm cost}(x; x^{(r_k)})$
for any $x \in {\rm lev}_{\rm init}$.
When $k \rightarrow \infty$,
$u_{\rm cost}(x^*; x^*) \le u_{\rm cost}(x; x^*)$.
Since $u_{\rm cost}(x; x^*)$ is a differentiable
convex function on $x$,
$0 \in \partial u_{\rm cost}(x^*; x^*) = \{\nabla u_{\rm cost}(x^*; x^*)\}$.
Then, $\nabla L_{\rm cost}(x^*) = \nabla u_{\rm cost}(x^*; x^*) = 0$.
Thus, $x^*$ is the unique point in the minimum set of $L_{\rm cost}$.
So, $(x^{(r_k)})_{k = 1,2,\cdots}$ is a sequence such that
$(L_{\rm cost}(x^{(r_k)}))_{k = 1,2,\cdots}$ converges to $\inf L_{\rm cost}$.
Because of the monotonicity of the MM sequence,
$(L_{\rm cost}(x^{(r)}))_{r = 1,2,\cdots}$ also converges to $\inf L_{\rm cost}$.
Thus, the MM sequence $(x^{(r)})_{r = 1,2,\cdots}$ converges to
the unique minimum point $x^*$
[\citet{Rockafellar_1970}, Corollary 27.2.2].



\begin{thebibliography}{}
%
%

%
%
 
\bibitem[Angelini et~al.(2016)]{Angelini+2016}
  Angelini,~L., et al. 2016, \procspie, 9905, 990514

\bibitem[Awaki et~al.(2014)]{Awaki+2014}
  Awaki,~H., et al. 2014, \ao, 53, 7664

\bibitem[Bishop(2006)]{Bishop_06}
  Bishop,~C.~M. 2006,
  Pattern Recognition and Machine Learning
  (New York: Springer New York)

\bibitem[Dempster, Laird, and Rubin(1977)]{Dempster+1977}
  Dempster,~A.~P., Laird,~N.~M., \& Rubin,~D.~B. 1977,
  Journal of the Royal Statistical Society, Series B (Methodological), 39, 1
  
\bibitem[Ducros et~al.(1970)]{Ducros+1970}
  Ducros,~G., Ducros,~R., Rocchia,~R., \& Tarrius,~A. 1970,
  \nat, 227, 152

\bibitem[Garnett(2023)]{Garnett_2023}
  Garnett,~R. 2023, Bayesian Optimization
  (Cambridge, UK: Cambridge University Press)
  
\bibitem[Hagino et~al.(2018)]{Hagino+2018}
  Hagino,~K., et al. 2018, Journal of Astronomical Telescopes,
  Instruments, and Systems, 4, 021409
  
\bibitem[Harrison et~al.(2013)]{Harrison+2013}
  Harrison,~F.~A., et al. 2013, \apj, 770, 103

  
\bibitem[Hastie, Tibshirani, and Wainwright(2015)]
  {Hastie_Tibshirani_Wainwright_2015}
  Hastie,~T., Tibshirani,~R., \& Wainwright,~M. 2015,
  Statistical Learning with Sparsity: The Lasso and Generalizations
  (London: Routledge)
  
\bibitem[Hitomi Collaboration et~al.(2018)]{Hitomi_Collaboration+2018}
  Hitomi Collaboration, et al. 2018, \pasj, 70, 15

  
\bibitem[Hunter and Lange(2000)]{Hunter_Lange_2000}
  Hunter,~D.~R., \& Lange,~K. 2000,
  Journal of Computational and Graphical Statistics, 9, 60
  
\bibitem[Kanamori et~al.(2016)]{Kanamori+2016}
  Kanamori,~T., et al. 2016,
  Continuous optimization for machine learning, (Tokyo: Kohdansha)
  (written in Japanease)

\bibitem[Kennel and Coroniti(1984)]{Kennel_Coroniti_1984}
  Kennel,~C.~F., \& Coroniti,~F.~V. 1984, \apj, 283, 694
   
\bibitem[Lucy(1974)]{Lucy_1974}
  Lucy,~L.~B. 1974, \aj, 79, 745
  
\bibitem[Madsen et~al.(2015)]{Madsen+2015}
  Madsen,~K.~K., et al. 2015, \apj, 801, 66

\bibitem[Matsumoto et~al.(2018)]{Matsumoto+2018}
  Matsumoto,~H., et al. 2018,
  Journal of Astronomical Telescopes, Instruments, and Systems, 4, 011212
  
\bibitem[Mori et~al.(2004)]{Mori+2004}
  Mori,~K., Burrows,~D.~N., Hester,~J.~J., Pavlov,~G.~G.,
  Shibata,~S., \& Tsunemi,~H. 2004, \apj, 609, 186
  
\bibitem[Morii, Ikeda, and Maeda(2019)]{Morii+2019}
  Morii,~M., Ikeda,~S., \& Maeda,~Y. 2019, \pasj, 71, 24
  
\bibitem[Nakazawa et~al.(2018)]{Nakazawa+2018}
  Nakazawa,~K., et al. 2018,
  Journal of Astronomical Telescopes, Instruments, and Systems, 4, 021410

\bibitem[Rees and Gunn(1974)]{Rees_Gunn_1974}
  Rees,~M.~J., \& Gunn,~J.~E. 1974, \mnras, 167, 1
  
\bibitem[Petre et~al.(1985)]{Petre+1985}
  Petre,~R., \& Serlemitsos,~P.~J. 1985, \ao, 24, 1833

\bibitem[Porth, Komissarov, and Keppens(2014)]{Porth_Komissarov_Keppens_2014}
  Porth,~O., Komissarov,~S.~S., \& Keppens,~R. 2014, \mnras, 438, 278

\bibitem[Pravdo, Angelini, and Harding(1997)]{Pravdo_Angelini_Harding_1997}
  Pravdo,~S.~H., Angelini,~L., \& Harding,~A.~K. 1997, \apj, 491, 808
  
\bibitem[Richardson(1972)]{Richardson_1972}
  Richardson,~W.~H. 1972,
  Journal of the Optical Society of America, 62, 55

\bibitem[Rockafellar(1970)]{Rockafellar_1970}
  Rockafellar,~R.~T. 1970, Convex Analysis
  (Princeton, New Jersey: Princeton University Press)
  
\bibitem[Takahashi et~al.(2016)]{Takahashi+2016}
  Takahashi,~T., et al. 2016, \procspie, 9905, 99050U

\bibitem[Varadhan and Roland(2008)]{Varadhan_Roland_2008}
  Varadhan~R., \& Roland,~C. 2008
  Scandinavian Journal of Statistics, 35, 335
  
\bibitem[Weisskopf et~al.(2000)]{Weisskopf+2000}
  Weisskopf,~M.~C., et al. 2000, \apjl, 536, L81

\bibitem[Zhou, Alexander, and Lange(2011)]{Zhou_Alexander_Lange_2011}
  Zhou,~H., Alexander,~D., \& Lange,~K. 2011,
  Statistics and Computing, 21, 261
\end{thebibliography}
\end{document}